\documentclass[prb,showpacs,amsmath,amssymb,a4paper]{revtex4}% Physical Review B

\usepackage{graphicx}% Include figure files
\usepackage{dcolumn}% Align table columns on decimal point
\usepackage{bm}% bold math
\usepackage{color}% color notes

\begin{document}

%\preprint{}

%==============================================================================
\title{Quantum phase diagram of the integrable $p_x+ i p_y$ fermionic superfluid}
%==============================================================================

\author{Stefan M.A. Rombouts}
 \email{rombouts@iem.cfmac.csic.es}
\author{Jorge Dukelsky}
\affiliation{%
  Instituto de Estructura de la Materia,
  C.S.I.C.,
  Serrano 123, E-28006 Madrid, Spain}%
\author{Gerardo Ortiz}
\affiliation{%
  Department of Physics,
  Indiana University,
  Bloomington IN 47405, USA
}%

\date{\today}% It is always \today, today,

\begin{abstract}
We determine the zero temperature quantum phase diagram of a $p_x+ip_y$ pairing model
based on the exactly solvable hyperbolic Richardson-Gaudin model.
We present analytical and large-scale numerical results for this model.
In the continuum limit, the exact solution exhibits
a third-order quantum phase transition,
separating a strong-pairing from a weak-pairing phase.
The mean field solution allows to connect these results
to other models with $p_x+ip_y$ pairing order.
We define an experimentally accessible characteristic length scale,
associated with the size of the Cooper pairs,
that diverges at the transition point,
indicating that the phase transition is of a confinement-deconfinement type
without local order parameter.
We show that this phase transition is not limited to the $p_x+ip_y$ pairing model,
but can be found in any representation of the hyperbolic Richardson-Gaudin model
and is related to a symmetry that is absent in the rational Richardson-Gaudin model.
\end{abstract}

% PACS, the Physics and Astronomy Classification Scheme.
\maketitle

%==============================================================================
\section{Introduction}
%==============================================================================
\label{sec:intro}
One of the striking features of degenerate Fermi gases
whose constituents interact through an attractive potential
is that they can exhibit superfluidity or superconductivity.
The constituents may be neutral atoms, as is the case in trapped
Fermi gases  of $^{40}$K and $^6$Li \cite{Giorgini08} or superfluid $^3$He \cite{Legget1975},
or charged electrons as in conventional metallic superconductors, nucleons in heavy nuclei or nuclear matter \cite{Dean2003}.
The particular nature and symmetry of the attractive pairing interaction
may lead to exotic superfluid phases with complex order parameters,
as is well-known in the case of liquid $^3$He.
For order parameters with certain rotational symmetries,
a change in the coupling strength may induce a quantum phase transition (QPT)
separating different kinds of superfluid phases.
In a Fermi system whose pairing interaction has an $s$-wave character,
it is well understood that by increasing the coupling strength
there is a crossover, and not a QPT,
between a weak coupling Bardeen-Cooper-Schrieffer \cite{Bardeen57} (BCS)
and a Bose-Einstein condensate (BEC) phase \cite{Leggett1980}.
However, for higher rotational order,
the ground state can exhibit a QPT between qualitatively different superfluid states,
with a corresponding non-analyticity in the ground state energy.
In the present paper we focus on $p$-wave pairing
that might be encountered in ultracold Fermi gases
\cite{Regal2003,Zhang2004,Cheng2005,Zhang2008,Nishida2009,Cooper2009,Massignan2010}
or in exotic superconductors such as Sr$_2$RuO$_4$ \cite{Xia2006}.

As a schematic model for such systems we study the quantum phase diagram
of a spinless Fermi gas with $p_x+ip_y$ pairing interaction symmetry
\begin{equation}
  H_{p_x+ip_y} = \sum_{k} \frac{k^2}{2} c^{\dagger}_k  c^{\;}_k
        -  \sum_{k,k'}  V_{kk'} c^{\dagger}_k c^{\dagger}_{-k}  c^{\;}_{-k'} c^{\;}_{k'},
  \label{eq:hamhyperbolic}
\end{equation}
where $V_{kk'}=g_k^* g^{\;}_{k'}$ is a separable interaction,
with $g_k$ representing a complex function of wavevector $k$ and symmetry $p_x+ip_y$.
Though schematic,
this model captures the basic physics of the BEC and BCS sides of the phase diagram
and offers detailed insights into the phase transition mechanism.
Previous studies concentrated on the standard mean field description \cite{Read2000,Gurarie2005,Botelho2005,Iskin2006}.
The interesting point of the Hamiltonian in Eq.(\ref{eq:hamhyperbolic})
is that under certain conditions on the pairing coefficients $g_k$
the model is integrable and can be solved exactly with a Bethe ansatz \cite{Ibanez2009}.
This has the important consequence that it demonstrates that the BEC-BCS QPT
in this model is not an artefact of the mean field approximation,
even though the ground state correlation functions and energy
coincide with the mean field ones in the thermodynamic limit for attractive interactions.
Furthermore, the defining properties of the Bethe ansatz wave function
are so qualitatively distinct from its mean field counterpart that it deserves further investigation.
Indeed, as we will see, some features and insights resulting from the Bethe ansatz state
are absent in the mean field solution.

\begin{figure}[h!]
\includegraphics[width=16cm]{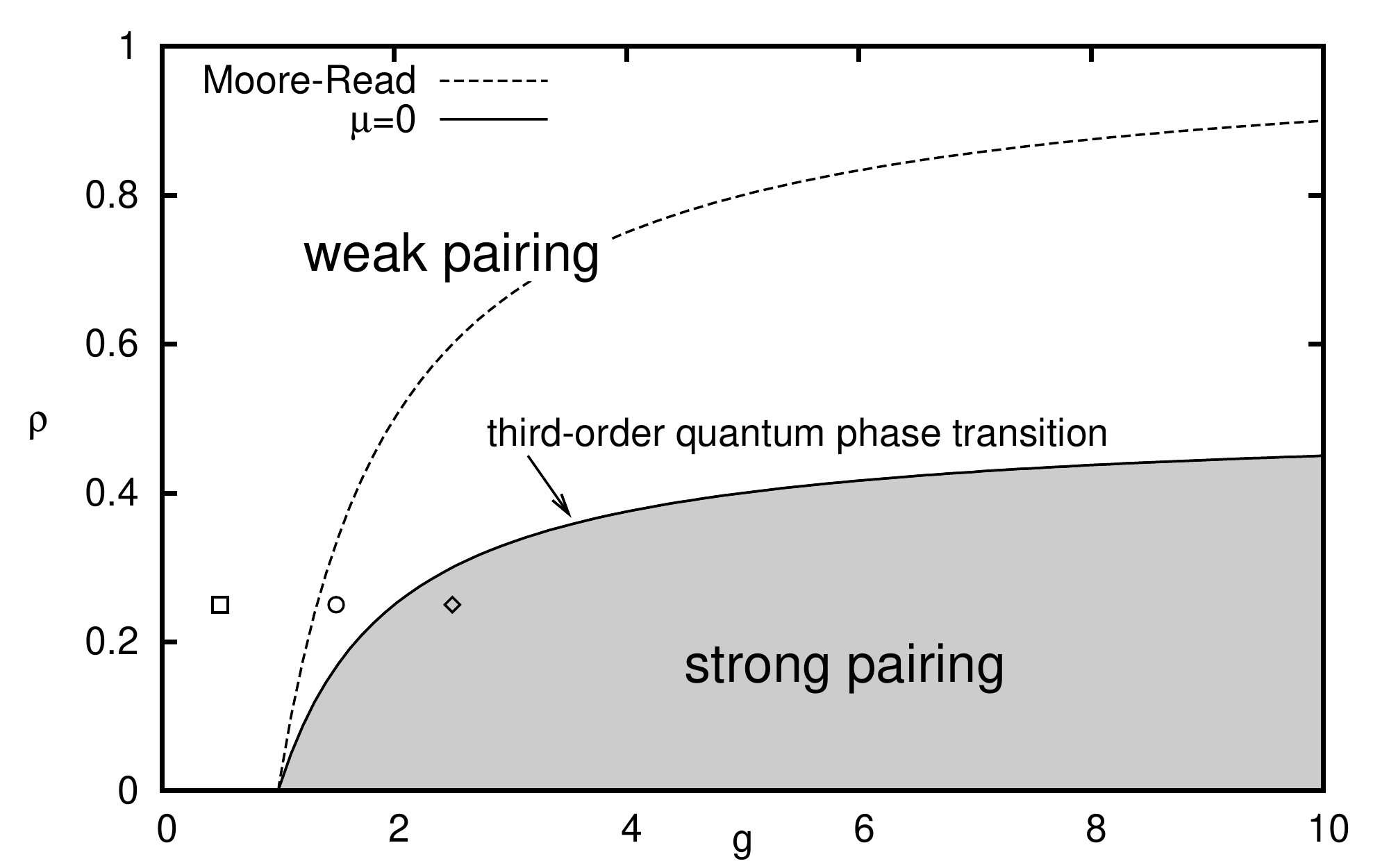}
\caption{\label{fig:hyperbolic_phasediagram}
Quantum phase diagram of the $p_x+ip_y$ or hyperbolic model
in terms of the fermion density $\rho$ and the attractive coupling strength $g$.
It shows two superfluid phases, a strong-pairing (confined) and a weak-pairing (deconfined) phase,
separated by a third-order confinement-deconfinement QPT at vanishing chemical potential, $\mu=0$.
Cooper pairs deconfine between the phase transition line
and the Moore-Read line (indicated by a dashed line).
The latter line corresponds to $\rho=1-1/g$ and vanishing total energy, $E=0$,
and represents a situation where all Cooper pairs are deconfined.
The small symbols at $\rho=0.25$ (quarter-filling)  indicate the configurations
that are displayed in detail in Figs.~\ref{fig:pairons126} and \ref{fig:densities126}.
}
\end{figure}

In this paper we restrict specific calculations to the two-dimensional attractive case
where a QPT with unusual characteristics signals the transition between two gapped superfluid phases,
one topologically non-trivial, known as weak pairing\cite{footnote1},
and another one characterized by tightly bound quasi-molecules, known as strong pairing \cite{Read2000}.
The quantum phase diagram of the two-dimensional $p_x+ip_y$ spinless fermion pairing Hamiltonian
displays several peculiarities (see Fig.~\ref{fig:hyperbolic_phasediagram}),
depending on the fermion density $\rho$.
For dense gases, $\rho > 1/2$, there is no QPT of any sort.
When $\rho < 1/2$, the model displays a QPT between a weak-pairing
and a strong-pairing phase,
while otherwise the ground state energy is an analytic function of the coupling strength $g$.
Volovik anticipated \cite{Volovik1985}, using an equivalent effective field theory,
that a QPT of this kind can be signaled by a discontinuous change of a topological quantum number.
We show that the transition is continuous and {\it third order}
(meaning that the third-order derivative of the ground state energy is discontinuous),
contrary to what has been claimed in the literature \cite{Nayak2008,Dunning2010}.
Furthermore, we demonstrate that this phase transition
can be related to an experimentally accessible length scale
that diverges logarithmically at the phase transition point.
We give an interpretation of this length scale
in terms of the size of the Cooper pairs,
with strong pairing corresponding to confined pairs
that start to deconfine when the system enters the weak-pairing phase,
and that eventually convert to plane waves in the weakly interacting limit.

A peculiar boundary in the quantum phase diagram
is the so-called Moore-Read line \cite{Moore1991} (shown as a dashed line in Fig.~\ref{fig:hyperbolic_phasediagram}).
The latter separates in principle a weak-coupling BCS regime from the weak-pairing phase
and is defined for a particular relation
between the density $\rho$ and the interaction strength $g$:
$\rho=1-1/g$.
At this line all Cooper pairs become deconfined.
As explained in detail in the text, our calculations indicate that there is no QPT
at the Moore-Read line, contrary to what is advocated in Refs.~\onlinecite{Ibanez2009,Dunning2010}.
Thus, the weak coupling BCS region and the weak-pairing phase are adiabatically connected.

Technically, we solve the more general hyperbolic (or XXZ) Richardson-Gaudin (RG) model \cite{Ortiz2005},
a particular realization of which is the $p_x+ip_y$ two-dimensional superfluid.
We present results for finite systems and,
for determining the quantum phase diagram, the relevant thermodynamic limit.
Calculating those results for the hyperbolic RG model
implies solving the corresponding Bethe ansatz equations,
a task that may happen to be technically challenging, specifically for large finite systems.
We present a new technique that solves the Bethe ansatz equations numerically in polynomial time.
For very large systems one can define a thermodynamic limit
that allows one to solve the equations for the integrable $p_x+ip_y$ model analytically \cite{Dunning2010}.

We want to emphasize that {\it any} hyperbolic RG model in the continuum limit
displays the non-analytic behavior observed at the equivalent $\mu=0$ line
as long as the parameters defining the model vanish at a given zero mode (e.g., $k=0$ mode).
The underlying algebraic structure of the hyperbolic RG model
allows us to prove a theorem that relates subspaces
with different number of pairs but with the same eigenvalues,
thereby generalizing a relation from Ref.~\onlinecite{Ibanez2009}
to all representations of the hyperbolic RG model.
These relations are inexistent in the case of the rational RG model
(e.g., the $s$-wave  BCS superconductor)
and are indeed at the root of the existence of the QPT between the weak- and strong-pairing phases.

In Sec.~\ref{sec:rg} we discuss the algebraic properties of the hyperbolic RG model.
In Sec.~\ref{sec:results} we present numerical and analytical results for the $p_x+i p_y$ model,
analyze the phase diagram and interprete the QPT mechanism.
Using an affine Lie algebra, we show in Sec.~\ref{sec:symmetry} that the hyperbolic
RG model possesses a symmetry that relates the two phases.
In the Appendices we give a detailed solution of the hyperbolic RG model for one pair (Appendix~\ref{app:onepair}),
for one level (Appendix~\ref{app:onelevel}),
in the continuum limit (Appendix~\ref{app:continuum})
and in the mean field approximation (Appendix~\ref{app:meanfield}).

%==============================================================================
\section{Algebraic formulation of the integrable $p_x+i p_y$ model
         and its solutions}
%==============================================================================
\label{sec:rg}

%==============================================================================
\subsection{The generalized Gaudin algebra}
%==============================================================================
\label{sec:integrable}

The $p_x+i p_y$ integrable pairing model can be obtained as a particular parameterization of the
exactly solvable hyperbolic family derived from the generalized Gaudin algebra.
We recapitulate here the main ingredients of the procedure to construct such exactly solvable models
that we have laid out in previous work~\cite{Ortiz2005}.
At the same time, this serves to demonstrate
that particular features of the zero temperature phase diagram of the $p_x + i p_y$ model
are not limited to this representation
and analogs can be found in many other representations of the same underlying Gaudin algebra.
Assuming a system characterized by $l$ levels (representations) of an $su(2)$ algebra labeled by the
(in principle arbitrary) quantum number $i$,
the generic form of the exactly solvable RG Hamiltonians is:
\begin{eqnarray}
  H &=& \sum_{i} \epsilon_i S^z_i
        - \sum_{i,j} (\epsilon_i-\epsilon_j) X(\eta_i,\eta_j) S^+_i S^-_j
        - \sum_{i,j} (\epsilon_i-\epsilon_j) Z(\eta_i,\eta_j) S^z_i S^z_j,
  \label{eq:rg_hamiltonian}
\end{eqnarray}
where the operators $S^+_i$, $S^-_i$ and $S^z_i$ close the $su(2)$ commutator algebra,
and $X(x,y)$ and $Z(x,y)$ are antisymmetric functions
depending on $l$ parameters  $\eta_i$ to be determined later on.
The generic Hamiltonian (\ref{eq:rg_hamiltonian}) commutes
with the squared spin operator of each level,
${\bf S}_i^2=S^z_i(S^z_i-1) + S^+_i S^-_i$,
and with the total spin operator ${\bf S}= \sum_i {\bf S}_i$.
Let us introduce the notation $s_i$
for the spin value of level $i$,
such that $\langle {\bf S}_i^2 \rangle = s_i(s_i+1)$.
Apart from this, one can evaluate the commutator between
two such Hamiltonians, $H$ and $H'$,
based on the same set of parameters $\eta_i$
but with distinct sets of parameters $\epsilon_i$ and $\epsilon_i'$.
It turns out that $H$ and $H'$ commute
provided that the functions $X(x,y)$ and $Z(x,y)$ fulfill
the following condition for all $x$, $y$ and $z$ known as the Gaudin condition \cite{Gaudin1976}:
\begin{equation}
  \left[ Z(x,y)-Z(x,z) \right] X(y,z) - X(x,y) X(x,z) = 0.
  \label{eq:gaudineq}
\end{equation}
A consequence of this condition is that
\begin{equation}
  X(x,y)^2 - Z(x,y)^2 = \Gamma,
  \label{eq:gaudingamma}
\end{equation}
where $\Gamma$ is a constant independent of $x$ and $y$.
Then, by a suitable choice of the parameters $\epsilon_i$ in Eq.(\ref{eq:rg_hamiltonian}),
one can define a set of $l$ linearly independent constants of motion,
which commute among themselves:
\begin{equation}
  R_i = S^z_i
        - \sum_{j,j\neq i} X(\eta_i,\eta_j) \left( S^+_i S^-_j + S^-_i S^+_j \right)
        - 2 \sum_{j,j\neq i} Z(\eta_i,\eta_j) S^z_i S^z_j.
  \label{eq:integrals_of_motion}
\end{equation}
Therefore, Hamiltonians of the form of Eq.(\ref{eq:rg_hamiltonian}) are integrable \cite{Ortiz2005},
and furthermore, they can be solved exactly by a Bethe ansatz.

It is convenient to define the following operators:
\begin{equation}
  {\sf S}^z(x) = -\frac{1}{2} - \sum_j Z(x,\eta_j) S^z_j,
 \ \ \ \
  {\sf S}^\pm(x) = \sum_j X(x,\eta_j) S^{\pm}_j.
\end{equation}
They fulfill the commutation relations of an {\em XXZ Gaudin algebra},
\begin{eqnarray}
  \left[ {\sf S}^z(x), {\sf S}^\pm(y) \right]  &=& \pm \left( X(x,y) {\sf S}^\pm(x) -Z(x,y) {\sf S}^z(y) \right),
 \nonumber \\
  \left[ {\sf S}^+(x), {\sf S}^-(y) \right]  &=& 2 X(x,y) \left( {\sf S}^z(x) - {\sf S}^z(y) \right).
 \label{eq:gaudinalgebra}
\end{eqnarray}
With this notation, the Bethe ansatz for the eigenstates of order $M$ takes the form
\begin{equation}
 | \Phi_M \rangle = \left(\prod_{\alpha=1}^{M} {\sf S}^{+}(E_\alpha) \right) | \nu \rangle.
 \label{eq:bethe_ansatz}
\end{equation}
where $| \nu \rangle$ is a vacuum state
such that $S^-_i| \nu \rangle=0$, and $S^z_i| \nu \rangle=-s_i | \nu \rangle$ for all $i$.
The unknowns $E_\alpha$ are solutions
of a set of nonlinear equations,
the RG equations:
\begin{equation}
  \sum_{i} s_i Z(\eta_i,E_\alpha)
      - \sum_{\alpha', \alpha' \neq \alpha} Z(E_{\alpha'},E_\alpha)  = \frac{1}{2},
         \ \ \forall \alpha.
 \label{eq:eigeneq}
\end{equation}
We will call the variables $E_\alpha$ {\em pairons} \cite{Dukelsky2002}.
Equation (\ref{eq:eigeneq}) does not depend on the parameters $\epsilon_i$,
which means that the eigenstates given by Eq.(\ref{eq:bethe_ansatz})
do not depend on these parameters either.
This is due to the fact that the ansatz from Eq.(\ref{eq:bethe_ansatz})
represents the eigenstates of the complete set of the integrals of motion $R_i$,
and therefore of any linear combination of $R_i$'s,
as given by the Hamiltonian (\ref{eq:rg_hamiltonian})
whose eigenvalues are given by
\begin{equation}
 E(\Phi_M)= \langle \nu | H | \nu \rangle
         + 2 \sum_{i,\alpha} s_i \epsilon_i Z(\eta_i,E_\alpha).
 \label{eq:eigenenergy}
\end{equation}
For $\Gamma=0$ one recovers the {\em rational} RG model,
that can be parameterized as
\begin{equation}
   X(x,y) = Z(x,y)  =  \frac{G}{x-y}
   \label{eq:rational}
\end{equation}
(any other expression for $X$ and $Z$
that fulfills Eqs.(\ref{eq:gaudineq},\ref{eq:gaudingamma}) at $\Gamma=0$
can be transformed to this form through a redefinition of parameters).
In this work we will focus on the properties
of the {\em hyperbolic} RG model, which correspond to $\Gamma <0$.
%This class derives its name from the parameterization proposed by Gaudin,
%$ Z(x,y)= G \coth(x-y) $, $X(x,y) = G/\sinh(x-y)$.
Any set of functions $X(x,y)$ and $Z(x,y)$ that fulfills
Eqs.(\ref{eq:gaudineq},\ref{eq:gaudingamma}) with $\Gamma=-\gamma^2$,
can be mapped onto the following parameterization \cite{Ortiz2005}:
\begin{equation}
 X(x,y) = 2\gamma \frac{\sqrt{x}\sqrt{y}}{x-y}, \ \ \ 
 Z(x,y) = \gamma \frac{x + y}{x - y}.
\end{equation}
We can simplify the discussion by choosing $\epsilon_i = \lambda \eta_i$,
with $\lambda$ a parameter that will be fixed later on.
Using this parameterization, and subtracting a diagonal term
$ 2\gamma \sum_i \eta_i {\bf S}^2_i$,
one obtains an interesting form for the Hamiltonian of Eq.(\ref{eq:rg_hamiltonian}):
\begin{eqnarray}
  H &=&  \lambda (1+2\gamma) \sum_{i} \eta_i S^z_i
        -  \lambda \gamma \sum_{i,j} (\eta_i+\eta_j) S^z_i S^z_j
        -  2\lambda \gamma \sum_{i,j} \sqrt{\eta_i\eta_j} S^+_i S^-_j.
  \nonumber \\
   &=&  \lambda \left( 1+2\gamma-2\gamma S^z \right) \sum_{i} \eta_i S^z_i
        -  2\lambda \gamma \sum_{i,j} \sqrt{\eta_i\eta_j} S^+_i S^-_j.
   \label{eq:sp_hamiltonian1}
\end{eqnarray}
Now we can take advantage of the fact that $S^z=\sum_iS^z_i$ is a constant of motion,
\begin{equation}
  S^z|\Phi_M \rangle = (M - L/2) | \Phi_M \rangle,
\end{equation}
where $L = 2 \sum_i s_i$ is the maximum value that the model allows for the order $M$
(therefore $L$ is a useful measure for the dimension of the model space;
e.g. if $s_i =1/2$ for all levels, then $L$ equals the number of levels $l$).
This allows us to fix the parameters $\lambda$ and $\gamma$ such that
\begin{equation}
 \lambda \left( 1+\gamma (L-2M+2) \right) =1,
  \ \ 2\lambda \gamma = G,
\end{equation}
leading to a separable Hamiltonian for the hyperbolic RG model:
\begin{equation}
  H_h =  \sum_{i} \eta_i S^z_i
        - G \sum_{i,j} \sqrt{\eta_i\eta_j} S^+_i S^-_j.
   \label{eq:sp_hamiltonian}
\end{equation}
The corresponding RG equations, Eqs.(\ref{eq:eigeneq}), reduce to:
\begin{equation}
  \sum_{i} \frac{s_i}{\eta_i - E_\alpha}
      - \sum_{\alpha', \alpha' \neq \alpha}\frac{1}{E_{\alpha'}-E_{\alpha}}
         - \frac{Q}{E_\alpha}=0
         ,
         \ \ \forall \alpha,
 \label{eq:sp_eigeneq}
\end{equation}
with
\begin{equation}
    Q = \frac{1}{2G} - \frac{L}{2} + M - 1.
 \label{eq:q}
\end{equation}
Equation (\ref{eq:eigenenergy}) for the energy simplifies to:
\begin{equation}
 E(\Phi_M)= \langle \nu | H_h | \nu \rangle + \sum_{\alpha} E_\alpha.
 \label{eq:sp_eigenenergy}
\end{equation}
For the remaining discussion we will assume that $\langle \nu | H_h | \nu \rangle=0$,
which amounts to a simple shift in the energy scale, without loss of generality.

In the pairing representations each $su(2)$ copy is associated with a single particle level $i$.
$M$ is the number of active pairs.
The vacuum $|\nu\rangle$ is now defined by a set of seniorities,
 $|\nu\rangle$ = $|\nu_1,\nu_2,\ldots,\nu_l \rangle$,
where the {\em seniority} $\nu_i$ is the number of unpaired particles in level $i$
with single particle degeneracy $\Omega_i$,
such that $s_i=(\Omega_i-2\nu_i)/4$.

In two spatial dimensions, one can define a representation of the $su(2)$ algebra
in terms of spinless fermions in momentum space, $c^\dagger_{k}, c^{\;}_{k}$.
One obtains a level $\eta_k$ for each pair of states $(k,-k)$,
where the index $k$ now refers to the momentum in two dimensions
(in order to avoid double counting we select $k_x>0$ to label the levels).
Furthermore, one can include a phase factor in the definition of ${\bf S}_k$:
\begin{eqnarray}
  S^z_k = \frac{1}{2}\left( c^\dagger_k c^{\;}_k + c^\dagger_{-k} c^{\;}_{-k}  - 1 \right),
  \ \ S^+_k = \frac{k_x + i k_y}{|k|} c^\dagger_k c^\dagger_{-k},
  \ \ S^-_k = \frac{k_x - i k_y}{|k|} c^{\;}_{-k} c^{\;}_k.
\end{eqnarray}
By taking $\eta_k = k^2$, one obtains the $p_x + i p_y$ model
presented by Iba\~{n}ez et. al \cite{Ibanez2009}:
\begin{equation}
  H_{p_x+ip_y} =  \sum_{k,k_x>0} \frac{k^2}{2}  \left( c^\dagger_k c^{\;}_k + c^\dagger_{-k} c^{\;}_{-k} \right)
        - G %\sum_{\begin{array}{c}{k,k' \\ k_x>0, k'>0} \end{array}}
            \sum_{\substack{k, k_x>0, \\ k', k'_x>0}}
            (k_x + i k_y) (k'_x - i k'_y) c^\dagger_k c^\dagger_{-k} c^{\;}_{-k'} c^{\;}_{k'}.
   \label{eq:pip_hamiltonian}
\end{equation}
In the case of cold atom gases, the spinless fermions operators represent atoms in the same hyperfine state.

%==============================================================================
\subsection{Singularities in the Richardson-Gaudin equations}
%==============================================================================
\label{sec:singularity}

Just like in the rational case,
the RG equations become singular when
two or more pairons approach the same level \cite{Rombouts2004,Dominguez2006}.
Analyzing the residues in the equation,
one finds that the number of singular pairons
has to be equal to $2s_k+1$ for a singularity around level $\eta_k$.
However, around $E_\alpha=0$
the Eq.(\ref{eq:sp_eigeneq}) for the hyperbolic RG model
displays a peculiar singularity not present in the rational models.
Let us analyze what happens when $N$ of the pairons converge to zero energy,
and $M-N$ pairons do not.
For the pairons that do not converge to zero,
one finds that the equations correspond
to a state of order $M-N$ with all pairons distinct from zero.
In the equations for the $N$ singular pairons
the contributions of the non-singular pairons and the levels $\eta_k$
are of order one and become negligible compared to the singular terms.
Therefore we can replace the non-singular terms by a single level,
with an averaged level parameter $\eta$ given by
\begin{equation}
  \frac{L/2-(M-N)}{\eta} =
  \sum_{k} \frac{s_k}{\eta_k}
      - \sum_{\alpha', E_\alpha' \nrightarrow 0}\frac{1}{E_{\alpha'}}.
\end{equation}
Appendix~\ref{app:onelevel} gives the analytical solution of the one-level problem.
One finds that a necessary condition to have $N$ values converging to zero
is that $2Q+1=N$, or
\begin{equation}
  \frac{1}{G}=L-2M+N+1.
\end{equation}
We recognize two special cases:
\begin{description}
 \item[special case (i)] $N=0$:
        \\ no pairons can converge to zero for $\frac{1}{G} \leq L-2M+1$;
        we will see that this boundary coincides with the phase transition line in Fig.~\ref{fig:hyperbolic_phasediagram}.
 \item[special case (ii)] $N=M$:
        \\ all pairons can converge to zero for $\frac{1}{G} = L-M+1$;
        this situation determines the so called {\em Moore-Read} line
        \cite{Moore1991,Ibanez2009}
        in Fig.~\ref{fig:hyperbolic_phasediagram}.
\end{description}
Between these two regimes,
a fraction of the pairons can converge to zero at integer values of $G^{-1}$.

This brings us to a qualitative view of the quantum phase diagram
of Fig. \ref{fig:hyperbolic_phasediagram}:
on the left side of the diagram one finds the weak coupling BCS region,
where the pairons behave much like in the rational model:
they form arcs in the complex plane.
If the coupling strength is increased,
one reaches the Moore-Read line,
that coincides with special case (ii) where all pairons collapse to zero.
Increasing the interaction strength, still in weak pairing,
some of the pairons can collapse to zero (at integer values of $G^{-1}$).
Finally, for strong interactions and $M<L/2$,
one can pass special case (i)
and make the transition to the strong-pairing phase.
Here the ground state corresponds to a configuration
where all pairons are real and negative.

At the Moore-Read line all pairons $E_\alpha$ converge to zero,
which means that the total energy $E$ also goes to zero.
This coincides exactly with the mean field result.
The corresponding wave functions are not the same
because the mean field wave function breaks pair number symmetry,
and can be written as a superposition of exact zero-energy states for a range of numbers of pairs:
\begin{equation}
  | \psi_{\sf mf} \rangle \propto
           \exp \left( \Delta \sum_{k,k_x>0} \frac{1}{k_x -i k_y} c^\dagger_k c^\dagger_{-k} \right) |\nu \rangle,
\end{equation}
while the exact wave function is given by the projection
of $| \psi_{\sf mf} \rangle$ onto the $M$-pair sector,
\begin{equation}
  | \psi_{M} \rangle 
      \propto \left( \sum_{k,k_x>0} \frac{1}{k_x-i k_y} c^\dagger_k c^\dagger_{-k}  \right)^M |\nu \rangle.
\end{equation}
Just below the line, when $N=M-1=L-1/G-1$,
one finds that the energy of the exact eigenstate
is equal to the energy $E_p$ of the one-pair state calculated in Appendix~\ref{app:onepair},
because of the symmetry to be shown in section \ref{sec:symmetry}.
Increasing the number of pairs by one while maintaining $G$ fixed,
moves the state to the Moore-Read line where the total energy is zero.
This means that for a small increase in density, $\Delta M/L = 1/L$
(which becomes infinitesimal in the thermodynamic limit),
there is a finite jump in energy, $\Delta E = |E_p|$.
This is interpreted by Iba\~{n}ez et al.
as a signature of a {\em zero${}^{th}$ order QPT}.
Note however, that $E$ scales proportional to $\omega L$,
where $\omega$ is the kinetic energy cutoff as defined in App.~\ref{app:onepair}.
Therefore, the shift in energy $\Delta E$ resulting from
a shift in density $\Delta M/L$ is given by
\begin{equation}
  \Delta E
      =  \frac{dE}{dM} \Delta M
      =  \omega \frac{dE/(\omega L)}{d(M/L)} \Delta M,
\end{equation}
which is of order $\omega$ for $\Delta M=1$
because the derivative is free of scales.
This means that a jump in energy of size $|E_p|$
for a change in density of size $1/L$ is not a sign of a phase transition
but rather a consequence of the pathological scaling behavior
of the energy in this model.

%As a matter of fact, this reasoning does not depend on $E=0$,
%and hence the statement of Iba\~{n}ez et al. would imply
%that there is a zero${}^{th}$ order QPT
%at any value of the density. Of course this cannot be true!

%==============================================================================
\section{Quantum Phase Diagram of the integrable $p_x+ip_y$ model}
%==============================================================================
\label{sec:results}

%----------------------------------------------------------------
\subsection{The case of finite systems}
%----------------------------------------------------------------

We would like to stress here that in order to obtain the exact solution of the pairing eigenproblem
for $M$ pairs in a system of size $L$,
one has to diagonalize a Hamiltonian matrix of dimension $L!/(M!(M-L)!)$.
The integrability of the RG model reduces the exponential complexity of the problem
to that of solving a set of $M$ non-linear equations in $M$ unknowns, Eq.(\ref{eq:sp_eigeneq}),
which can be done in polynomial time.
This justifies labeling these models as {\em exactly solvable}.
For instance, below we show exact results for $L=504$ and $M=126$,
which corresponds to solving an eigenvalue problem of dimension $10^{122}$,
while in our case each eigenvector arises as a particular solution of the $126$ coupled non-linear RG equations.

Although easily written down,
actually solving Eqs.(\ref{eq:sp_eigeneq}) for more than a few pairs turns out to be a cumbersome task
due to singularities that occur when some of the pairons approach the origin or a certain level value $\eta_k$,
or if two or more pairons approach each other.
For the rational model with all $s_k=1/2$, where the singularities entail at most two pairons,
Richardson already proposed a practical solution \cite{Richardson1966},
by rewriting Eqs.(\ref{eq:sp_eigeneq}) in terms of the real part of the pairon values
and the square of the difference between two pairon values.
If singularities with three or more pairons occur, this trick no longer works.
More complicated changes of variables have been proposed
to remediate this problem for the rational model \cite{Rombouts2004,Dominguez2006},
but in the hyperbolic model we are confronted with a situation where {\em all} pairons can converge to zero
and where the change of variables can no longer be performed with sufficient accuracy.
Standard gradient methods to solve the Eqs.(\ref{eq:sp_eigeneq}) directly
run into problems because the Jacobian matrix becomes ill-conditioned
well before the pairons reach the singularities.

\begin{figure}[ht]
\includegraphics[width=12cm]{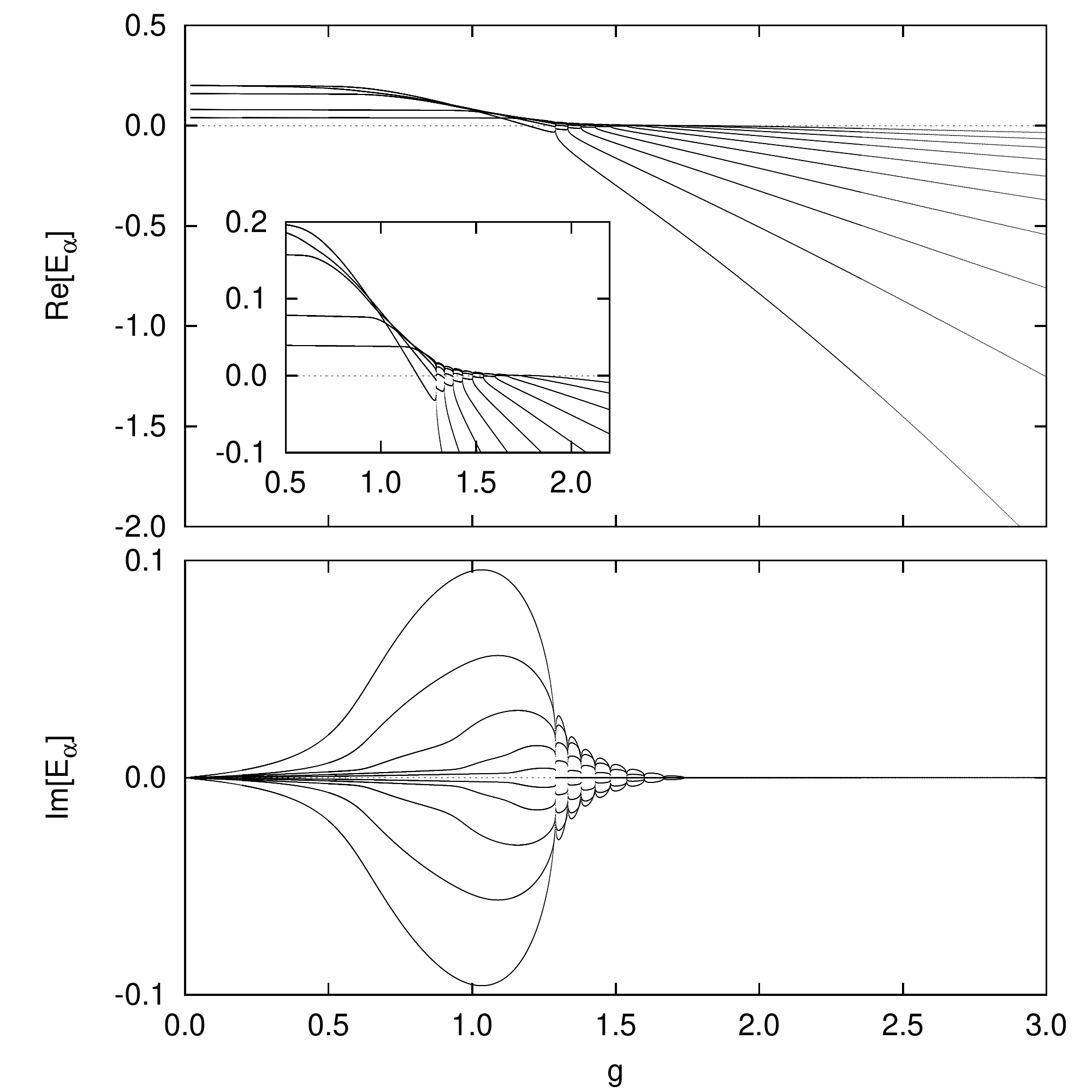}
\caption{\label{fig:pairons10} Real and imaginary parts of the spectral parameters
$E_\alpha$ as a function of $g=GL$ for 10 pairons in
a disk of dimension $L=40$ (quarter filling),
with levels as specified in Table \ref{tab:levels5}.
}
\end{figure}

\begin{figure}[ht]
\includegraphics[width=12cm]{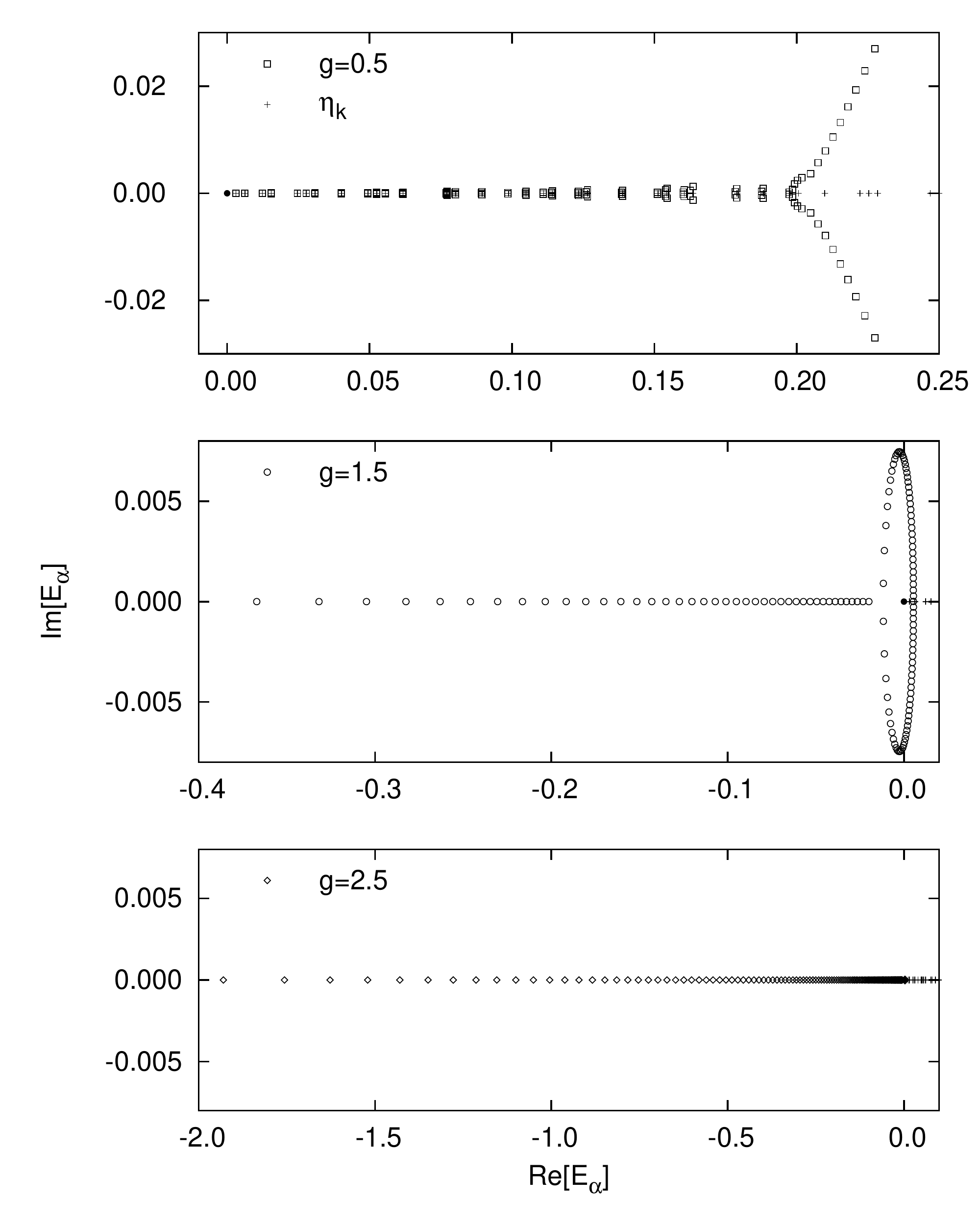}
\caption{\label{fig:pairons126} Pairon distribution for $L=504$ at quarter filling,
$\rho=0.25$, 
for $g=0.5$, $g=1.5$, and $g=2.5$.}
\end{figure}

\begin{figure}[ht]
\includegraphics[width=12cm]{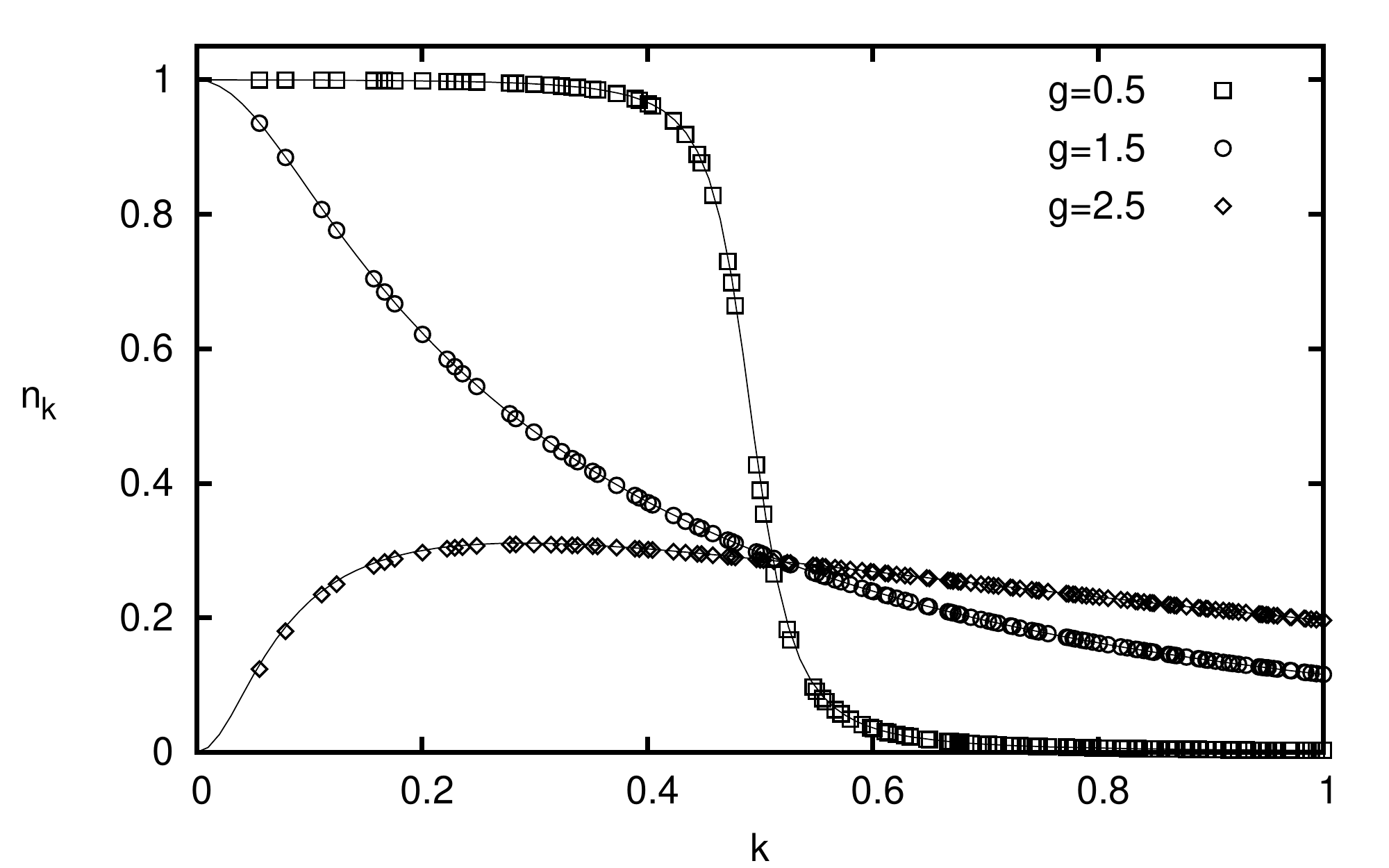}
\caption{\label{fig:densities126} Momentum density profiles, $n_k$, 
for the same parameter sets as in Fig.~\ref{fig:pairons126} (see symbols),
compared to the mean field solution for the same system (continuous lines).}
\end{figure}

\begin{figure}[h!]
\includegraphics[width=12cm]{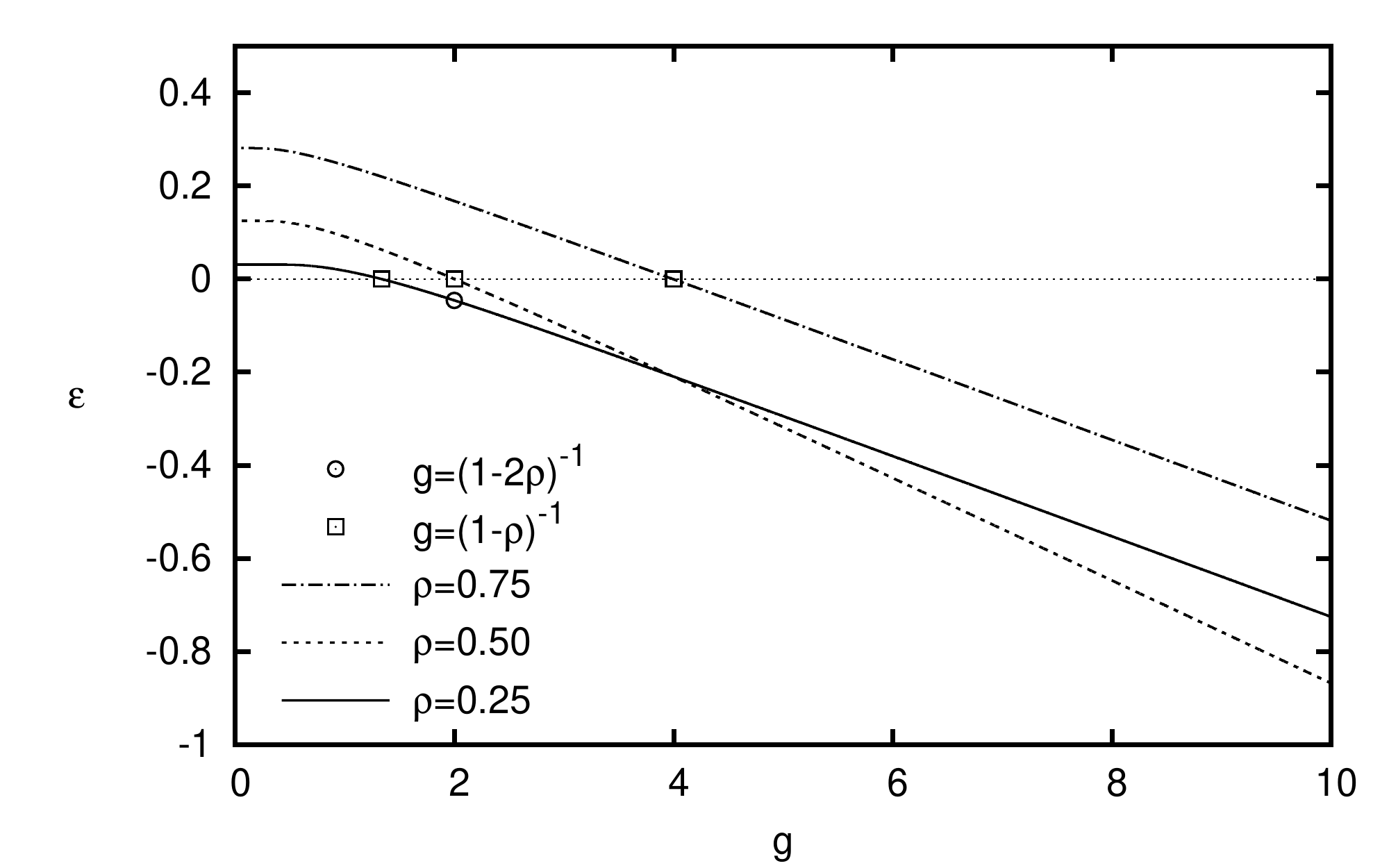}
\caption{\label{fig:energy} Analytical results for the energy density $\varepsilon$
of the integrable $p_x+ip_y$ model as a function of the coupling strength $g$ 
for representative densities $\rho$,  in the thermodynamic limit.
The symbol $\odot$ marks the QPT point, while $\boxdot$ the Moore-Read line.}
\end{figure}

Our numerical strategy to solve the RG equations
is to start from an educated guess for the position of the $M$ pairons in the weak coupling limit. 
In this limit the RG equations decouple into $M$ independent equations (one for each level).
As discussed in Appendix~\ref{app:onelevel},
the solution for each of these generalized Stieltjes equations
is given by the roots of a Jacobi polynomial,
which we use as a starting point for the iterative procedure at small values of $G$.
The value of the coupling strength $G$ is then gradually increased up to the desired value.
At each step we solve the Eqs.(\ref{eq:sp_eigeneq}) using a variation of the Levenberg-Marquardt algorithm\cite{Levenberg1944}, %,Marquardt1963},
where care has been taken to avoid singularities in the Jacobian matrix.
In order to avoid the typical divergencies that burden the numerical procedures in intermediate steps,
we modify the level parameters $\eta_k$ and the central charge $Q$ by adding a modulated artificial imaginary part.
Once the desired value of $G$ is obtained, a new iterative procedure is initiated,
in which the imaginary parts of $\eta_k$ and $Q$ are gradually reduced to obtain the final solution.
For the numerical results displayed in Figs.~\ref{fig:pairons10} to \ref{fig:radius},
we have taken the kinetic energy cutoff $\omega=1/2$,
such that the level parameters $\eta_k$ range from 0 to 1.

\begin{table}
\protect\begin{tabular}
[c]{c|c|c|c|c|c|c|c|c|c|c|c|c|c}
$\eta_{k}$ & $0.04$ & $0.08$ & $0.16$ & $0.20$ & $0.32$ & $0.36$ & $0.40$ & $0.52$ & $0.64$ & $0.68$
& $0.72$ & $0.80$ & $1.00$ \\
\hline
$s_{k}$ & 1 & 1 & 1 & 2 & 1 & 1 & 2 & 2 & 1 & 2 & 1 & 2 & 3
\protect\end{tabular}
\caption{Level parameters $\eta_k$ and $s_k$ for a disk with a radius of five unit cells in a two-dimensional square lattice. \label{tab:levels5}}
\end{table}

As a first illustration of the properties of the exact solution in the different regions of the quantum phase diagram,
we treat ten pairs in a two dimensional lattice in a disk geometry 
with a five unit cell radius.
The system has a total active space of dimension $L=2\sum_k s_k=40$,
hence this amounts to quarter filling, i.e. $\rho=0.25$.
The resulting level parameters are depicted in Table~\ref{tab:levels5}.
In the extreme weak coupling limit the ten pairs fill the lowest four levels.
With increasing coupling the pairons expand in the complex plane
lowering their real part, and consequently the total energy.
There is a particular value of the scaled coupling strength $g=GL$
at which all pairons collapse to zero (the Moore-Read line),
later on expelling one pairon with a negative real value.
{}From this point up to the QPT at $Q=0$ there are a series of collapses with nine, eight, etc. pairons,
and after each one a new real negative pairon is produced.
The exact solution for this small system shows the qualitative behavior of the pairons in the quantum phase diagram (Fig.~\ref{fig:hyperbolic_phasediagram}).
In order to get a more quantitative picture of the pairon distribution in the three regions of the quantum phase diagram,
we plot in Fig.~\ref{fig:pairons126} the pairon distributions for three representative values of the coupling strength, $g=0.5, 1.5, 2.5$,
at quarter filling for a disk of radius 18 corresponding to a total pair degeneracy $L=504$,
indicated in Fig.~\ref{fig:hyperbolic_phasediagram} by their respective symbols\cite{footnote2}.
One notes the distinctive features:
in weak coupling BCS part of the pairons stick to the lower part of the 
real positive axis, while the remaining pairons form a double arc.
Approaching the Moore-Read line it looks like the arc is going to close
around the origin,
but just at the Moore-Read line all the pairons collapse to zero.
A first real negative pairon emerges.
In the intermediate weak-pairing region
a successive series of collapses ensues,
at integer values of $Q$,
each time producing one more real negative pairon
and reducing the size of the arcs around the origin.
When the last pairon turns real and negative,
the system enters the strong-pairing phase.
From then on the most negative pairon diverges
proportional to the interaction strength $G$,
while the least negative pairon converges to a finite value
that can be related to the condensate fraction,
as we will see later on.

In Fig.~\ref{fig:densities126} we show the momentum density distributions
corresponding to each of the configurations in Fig.~\ref{fig:pairons126}.
In the weak-pairing region one sees that the lowest momentum states
are fully occupied.
On the other side of the phase transition
the lowest momentum states are vacated,
resulting in a non-trivial topology for the two-dimensional
momentum distribution.

%----------------------------------------------------------------
\subsection{The thermodynamic limit}
%----------------------------------------------------------------

\begin{figure}[ht]
\includegraphics[width=12cm]{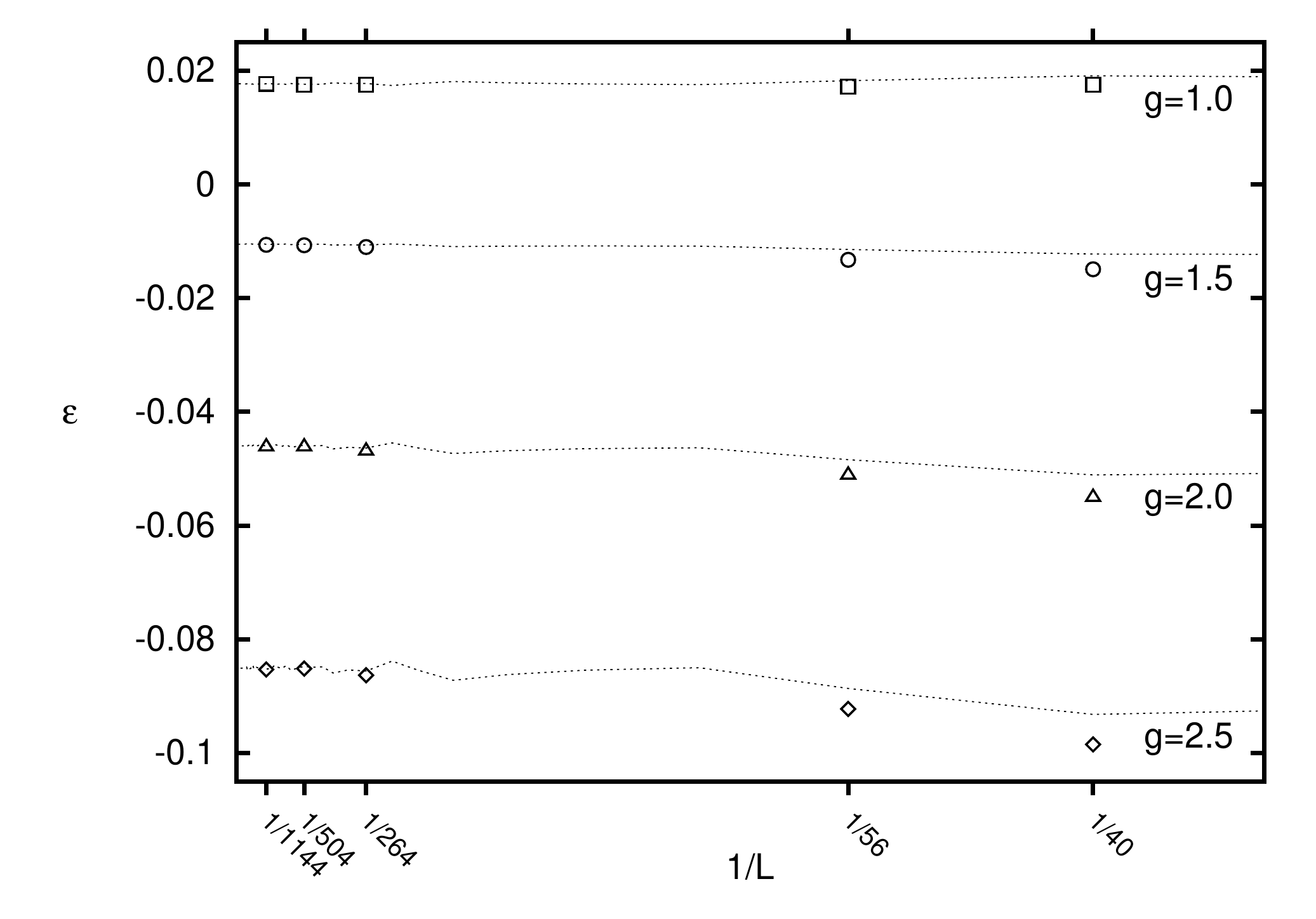}
\caption{\label{fig:thermolimit} Finite-size scaling of the energy
density $\varepsilon$ at quarter filling for several interaction strengths
(corresponding to distinct regions of the phase diagram),
both for the exact solution (symbols)
as for the mean field solution (lines).}
\end{figure}

We are interested in determining the quantum phase diagram of the integrable $p_x+ip_y$ model
defined by the Hamiltonian~(\ref{eq:pip_hamiltonian}).
To explore the possible QPTs
we have to properly define the thermodynamic limit for this model.
{} From Fig. \ref{fig:thermolimit} we see that for large systems
the exact solution and the mean field solution tend to the same results.
Particularly, in mean field theory one observes
that for static values of the chemical potential $\mu$ and the pairing field $\Delta$,
the quantities $M$, $E$ and $G^{-1}$ scale proportional with $L$.
We can describe the thermodynamic limit when $L$ goes to infinity
in terms of intensive quantities such as the scaled interaction strength $g=GL$
and the density $\rho=M/L$.
At the same time we have to establish a cutoff $\omega$ in kinetic energy,
such that the energy density $\varepsilon=E/L$ remains finite,
and a density of levels $\varrho(\eta)$, such that $\int_0^{2\omega} \varrho(\eta) d\eta=1$.
The cutoff $\omega$ determines the energy scale of the problem.
It can be renormalized by relating it to the energy of a single pair as explained in Appendix~\ref{app:onepair}.

\begin{figure}[ht]
\includegraphics[width=12cm]{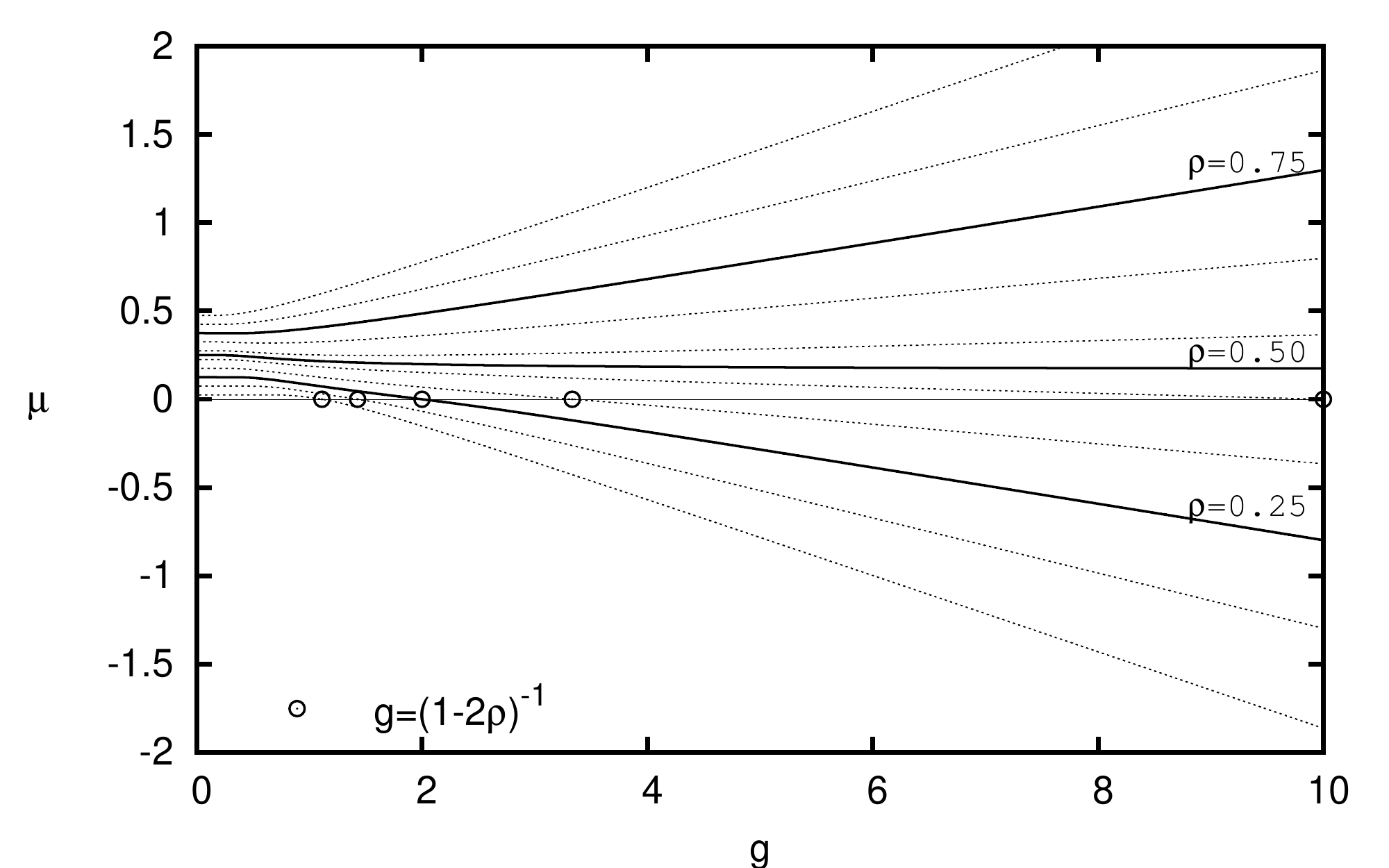}
\caption{\label{fig:mu} Analytical results for the chemical potential $\mu$ as a function of $g$,
for densities $\rho=0.05, 0.15,\ldots 0.95,$
in ascending order.
The symbols mark the transition points at $g=(1-2\rho)^{-1}$.}
\end{figure}

\begin{figure}[ht]
\includegraphics[width=12cm]{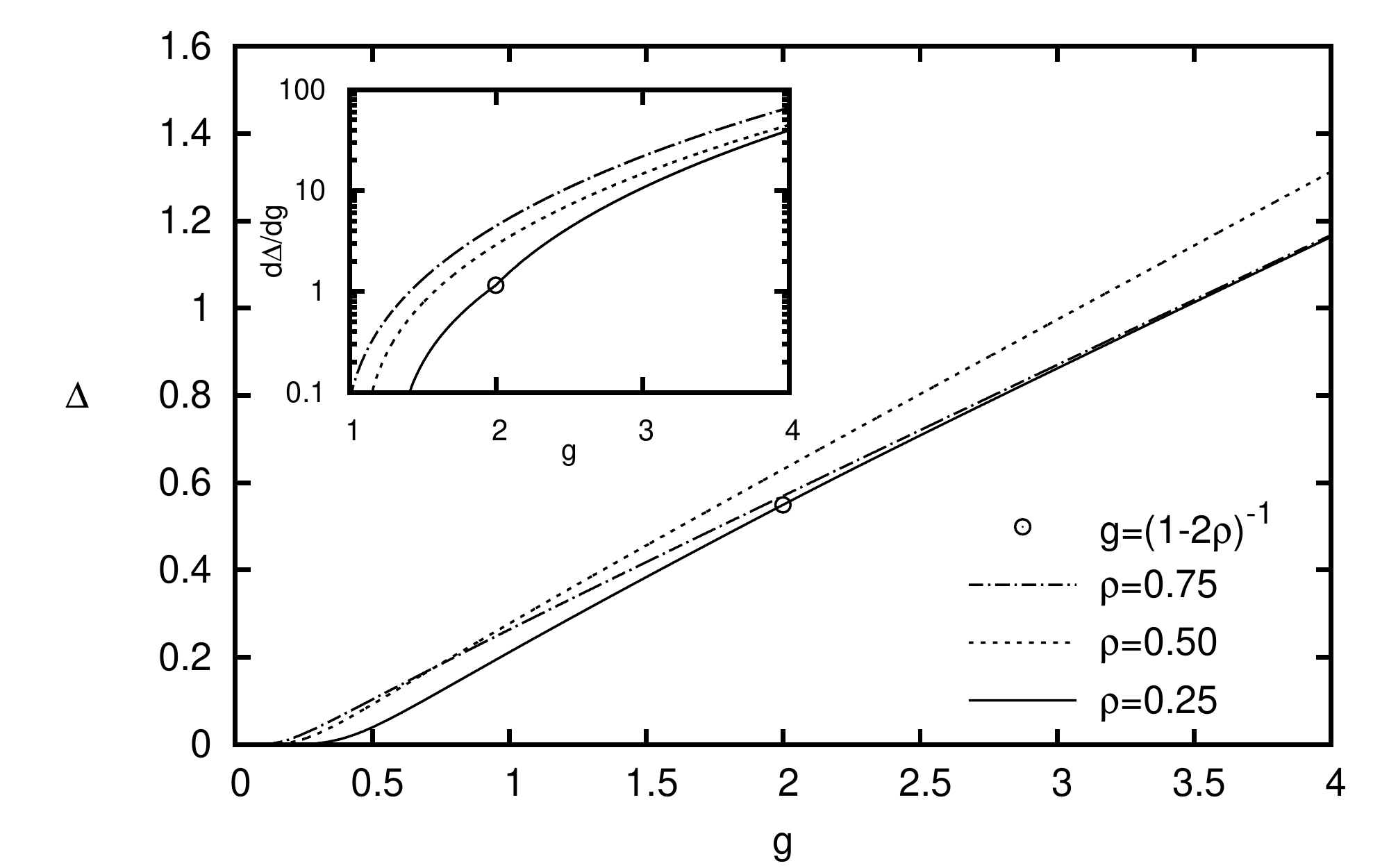}
\caption{\label{fig:delta} Analytical results for the pairing field $\Delta$
and its derivative $d\Delta/dg$ in the inset,
as a function of $g$ for various densities.}
\end{figure}

As demonstrated in Appendices~\ref{app:continuum} and \ref{app:meanfield},
the exact solution and the mean field solution coincide in the thermodynamic limit
and yield the following integral equations:
\begin{eqnarray}
   \int_0^{2\omega} \varrho(\eta) \frac{\eta}{\sqrt{(\eta-2\mu)^2+4\eta \Delta^2}} d\eta
       &=& \frac{1}{g},
  \label{eq:analytic_meanfield1}
  \\
   \int_0^{2\omega} \varrho(\eta) \frac{2\mu}{\sqrt{(\eta-2\mu)^2+4\eta\Delta^2}} d\eta
       &=& \frac{1}{g} -1+2\rho.
  \label{eq:analytic_meanfield2}
\end{eqnarray}
Note that the chemical potential $\mu$ and
the pairing field $\Delta^2$ scale proportional to $\omega$, and can be 
determined from the self-consistent solution of the above equations (see Figs 
\ref{fig:mu} and \ref{fig:delta}).
The energy density in the thermodynamic limit is given by
\begin{equation}
 \varepsilon =\frac{\omega}{2} + \mu (2\rho-1) + \frac{\Delta^2}{g}
      -  \frac{1}{2}\int_0^{2\omega} \varrho(\eta) \sqrt{ \left(\eta-2\mu\right)^2 + 4 \eta \Delta^2 } d\eta.
    \label{eq:analytic_energy1}
\end{equation}
One sees that Eqs.(\ref{eq:analytic_meanfield1},\ref{eq:analytic_meanfield2}) coincide
with the conditions
\begin{equation}
  \frac{d\varepsilon}{d\mu} = 0,
  \ \ \ \
  \frac{d\varepsilon}{d\Delta} = 0,
 \label{eq:analytical_ederivatives}
\end{equation}
meaning that the mean field solution minimizes the energy written
as a  function of the mean field parameters $\mu$ and $\Delta$.

For the two-dimensional $p_x+ip_y$ model one finds that $\varrho(\eta)=1/(2\omega)$,
and one can solve the integral appearing in Eq.(\ref{eq:analytic_energy1}) analytically:
\begin{eqnarray}
 \varepsilon &=& \frac{\omega}{2} + \mu (2\rho-1) + \frac{\Delta^2}{g}
          + \frac{|\mu|(\Delta^2-\mu)}{2\omega}
          - \frac{\omega+\Delta^2-\mu}{2\omega} \sqrt{(\omega-\mu)^2+2\omega\Delta^2}
    \nonumber \\ & &
              + \frac{\Delta^2 (\Delta^2-2\mu)}{2\omega}
                  \ln \left( \frac{\omega+\Delta^2-\mu+\sqrt{(\omega-\mu)^2+2\omega\Delta^2}}{\Delta^2-\mu+|\mu|} \right) .
    \label{eq:analytic_energy2}
\end{eqnarray}
{}From Eqs.(\ref{eq:analytical_ederivatives})
we obtain explicit expressions for $g$ and $\rho$:
\begin{eqnarray}
 g^{-1} &=& \frac{1}{\omega}
   \left[\sqrt{(\omega-\mu)^2+2 \omega \Delta^2} - |\mu|
         + \left( \mu - \Delta^2 \right)
        \ln\left(
      \frac{\omega-\mu+\Delta^2+\sqrt{(\omega-\mu)^2+2\omega \Delta^2}}
                    {\Delta^2 - \mu + |\mu|}\right)
    \right],
    \label{eq:analytic_g}
 \\
 \rho &=& \frac{1}{2} -\frac{1}{2\omega}
         \left[\sqrt{(\omega-\mu)^2+2 \omega \Delta^2} - |\mu|
           -\Delta^2  \ln\left(
      \frac{\omega-\mu+\Delta^2+\sqrt{(\omega-\mu)^2+2\omega\Delta^2}}
                    {\Delta^2 - \mu + |\mu|}\right) \right].
    \label{eq:analytic_rho}
\end{eqnarray}
This allows us to rewrite Eq.(\ref{eq:analytic_energy2}) in a more elegant form:
\begin{equation}
  \varepsilon
      =  \rho (\omega+\mu) + \frac{\Delta^2}{2g} - q \frac{\omega \Delta^2}{\mu} - \frac{\mu+|\mu|}{2},
    \label{eq:analytic_energy3}
\end{equation}
where
\begin{equation}
  q= Q/L = \frac{1}{2g} -\frac{1}{2} + \rho.
\end{equation}
The appearance of the absolute value of $\mu$ clearly indicates
that there exists a non-analyticity in $\varepsilon$ at $\mu=0$.

%----------------------------------------------------------------
\subsection{Derivatives of the energy density}
%----------------------------------------------------------------

The third term of Eq.(\ref{eq:analytic_energy3}) might suggest a singularity in $\varepsilon$ at $\mu=0$
and hence a zeroth order QPT at this point.
However, it turns out that $\mu=0$ coincides with $q=0$ such that the singularity in $\varepsilon$ is canceled out
and as a result $\varepsilon$ is a continuous function of $g$ and $\rho$
over the whole parameter range.
A special point that one observes in Eqs.(\ref{eq:analytic_g},\ref{eq:analytic_rho})
corresponds to $\mu=\Delta^2/2$, which leads to $g^{-1}=1-\rho$.
This is the {\em Moore-Read} line we discussed in Sec.~\ref{sec:singularity}.
As argued above and as can be seen from Fig.~\ref{fig:energy},
the energy vanishes at this point but is continuous.
Consequently, there is no zeroth order QPT in this model
(contrary to previous reports in the literature \cite{Ibanez2009,Dunning2010}).

The fourth term in Eq.(\ref{eq:analytic_energy3}) might suggest a discontinuity in the derivative $d\varepsilon/d\mu$ at $\mu=0$.
However, it turns out that the third term also has a discontinuity in its derivative,
that exactly cancels the one in the fourth term.
To formalize this, let us consider the derivatives of $\varepsilon$ with respect to $\rho$ and $g$:
\begin{equation}
  \frac{\partial\varepsilon}{\partial\rho} = 2\mu,
  \ \ \ \
  \frac{\partial\varepsilon}{\partial g} = - \Delta^2/g^2.
\end{equation}
In Eqs.(\ref{eq:analytic_g},\ref{eq:analytic_rho}) we see that $g^{-1}$ and $\rho$ are continuous functions
of $\mu$ and $\Delta$.
Hence, $\mu$ and $\Delta$ are continuous functions of $\rho$ and $g$,
and therefore the derivatives of $\varepsilon$ with respect to $\rho$ and $g$
are continuous functions of $\rho$ and $g$.
We conclude that there is no first order QPT in this model.
Note also that in the strong coupling limit
$\Delta$ tends to the same value for
$\rho=0.25$ and $\rho=0.75$ (see Fig. \ref{fig:delta}).
This is a consequence of the symmetry that we will discuss in Sec.~\ref{sec:symmetry}.

\begin{figure}[h!]
\includegraphics[width=12cm]{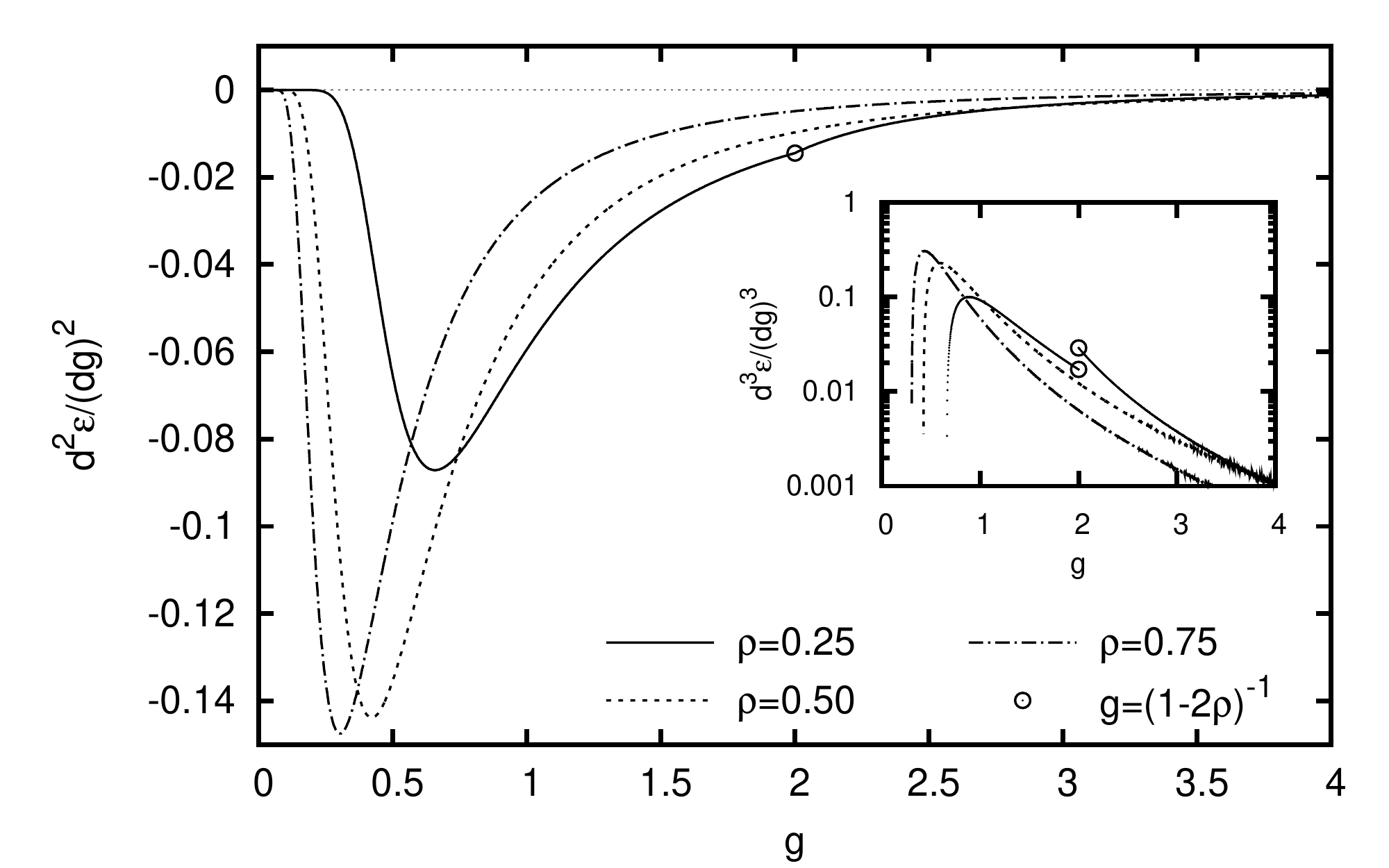}
\caption{\label{fig:ederivative}
  Higher-order derivatives of the energy density $\varepsilon$ as a function of $g$ for various densities.
  The small circles mark the transition point at $g=(1-2\rho)^{-1}$.
}
\end{figure}

For the higher order derivatives, we resort to a graphical representation.
{}From Fig.~\ref{fig:ederivative}
we see that there is a discontinuity in the third-order derivative of $\varepsilon$.
The discontinuity occurs when $\mu=0$.
If we set $\mu=0$ in Eqs.(\ref{eq:analytic_g},\ref{eq:analytic_rho}),
we find that this coincides with $g^{-1}=1-2\rho$.
This transition line has been identified in previous works as the Read-Green
line\cite{Read2000,Ibanez2009}.
We see that a third-order QPT occurs at this line,
for densities below half filling, i.e. $\rho\leq1/2$.

%-----------------------------------------------------------------------------
\subsection{Characterization of the superfluid phases}
%-----------------------------------------------------------------------------
In $s$-wave pairing models (rational RG model) the transition from a weak to a strong-pairing phase
is often described in terms of a ``BCS-BEC'' crossover,
where on the weak side one has dilute pairs moving in a coordinated way
as described by BCS theory,
while on the strong side the fermion pairs behave as molecules
that become perfect bosons in the strong coupling limit,
where they form a pure BEC.
In the $p_x+ip_y$ case the picture differs in an essential way:
we see that instead of a crossover
there is a sharp third-order QPT at $g^{-1}=1-2\rho$.

\begin{figure}[h!]
\includegraphics[width=12cm]{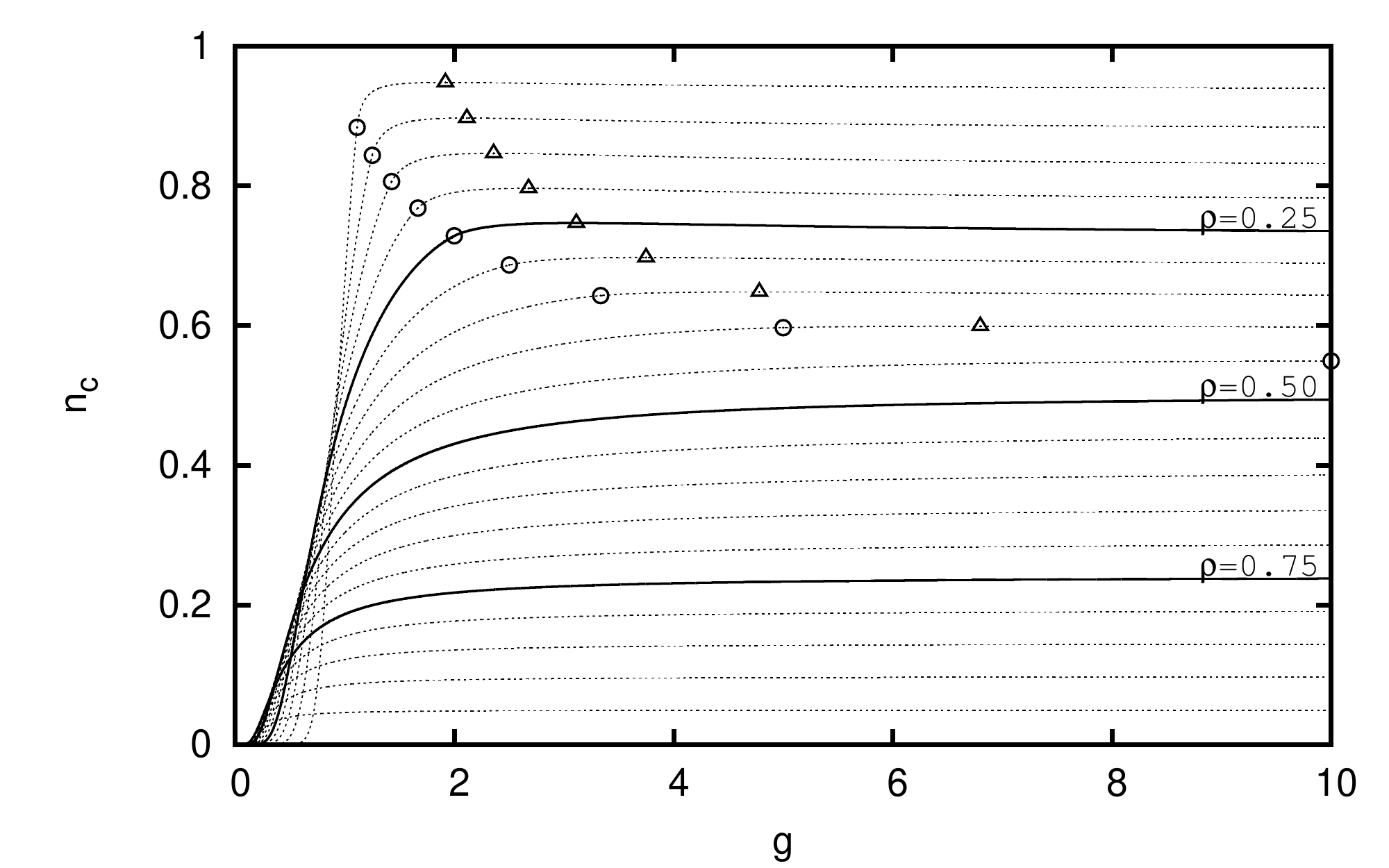}
\caption{ Condensate fraction of pairs, $n_c$,  as a function of $g$,
for densities $\rho=0.05, 0.10, \ldots, 0.95,$
in descending order.
The circles mark the transition points at $g=(1-2\rho)^{-1}$,
while the triangular symbols mark the point where $n_c$ is maximal.}
\label{fig:condfrac}
\end{figure}

Because the exact ground state correlation functions 
coincide with the mean field ones
in the thermodynamic limit,
we can use~Eq.(\ref{eq:mfsmink}) to
evaluate the condensate fraction $n_c$,
following the standard definition of Yang \cite{Yang1962}:
\begin{eqnarray}
  n_c &=& \frac{1}{M} \sum_{k,k_x>0} \left|\left\langle S^-_k \right\rangle \right|^2
 \nonumber \\
    & = & \frac{L \Delta^2}{2\omega M} \int_0^{2\omega} \frac{\eta}{(\eta-2\mu)^2+4\eta\Delta^2} d\eta
 \nonumber \\
   &=& \frac{\Delta^2}{2\omega \rho}
       \frac{a \ln\left(1-{2\omega}/{a}\right) - b \ln\left(1-{2\omega}/{b}\right)}
            {a-b},
 \label{eq:condfrac}
\end{eqnarray}
where  $a$ and $b$ are such that
$$
 (\eta-2\mu)^2 +4 \eta \Delta^2 = (\eta-a)(\eta-b).
$$
As explained in Appendix~\ref{app:continuum},
$a$ and $b$ are the lower and upper bounds
of the line of pairons in strong pairing in the thermodynamic limit.
We have used the mean field expectation values
because they coincide with the exact results in the thermodynamic limit.
For finite systems one could evaluate the exact correlation functions using techniques 
similar to the ones presented in Ref.~\onlinecite{Faribault2008} for the rational RG model.
Figure \ref{fig:condfrac} displays the results
for various densities.
One observes that a pair condensate develops at large interaction strength.
Two features draw attention:
first of all, we see that for a fixed density $\rho<1/2$
the condensate fraction reaches a maximum
in the strong-pairing phase at a finite value of $g$,
and then slightly decreases to its limiting value for $g\rightarrow \infty$.
This means that the fermion pairs do not behave as pure bosons,
but rather as bound molecules with a certain spatial extension,
such that the Pauli principle prohibits a complete condensation of all molecules
\cite{Rombouts2002}.
Secondly, note that because we study here the condensate fraction
as a function of the interaction strength,
the limiting value of $n_c$ in the strong BEC limit is not 1,
contrary to what one is accustomed to in three dimensions for the 
$s$-wave case \cite{Ortiz2005a},
where one evaluates $n_c$ as a  function of the ratio
between interparticle distance and scattering length.

To understand the behavior of the condensate fraction
in the strong coupling limit,
we evaluate $\mu$ and $\Delta^2$ for $g \gg 1$ up to leading order in $g$:
\begin{eqnarray}
 \mu & \sim & - 2\omega g (1-2\rho/3) (1-2\rho)/4,
 \\
 \Delta^2 & \sim & 2\omega g^2 (1-2\rho/3)^2 \rho/2.
% \\
% a & \sim & - 2\omega \frac{(1-2\rho)^2} {8\rho},
% \\
% b & \sim & - 4\omega g^2 \rho (3-2\rho)^2/9.
\end{eqnarray}
Using the expressions for $a$ and $b$ in terms of $\mu$ and $\Delta^2$
from  Appendix~\ref{app:continuum},
and substituting these in Eq.(\ref{eq:condfrac}),
one finds that in the strong coupling limit
\begin{equation}
  \lim_{L \rightarrow \infty, g \rightarrow \infty} n_c
   = \frac{1}{4\rho} \left[ 1 - \frac{(1 - 2 \rho)^2}{4 \rho} \ln\left( \frac{1+2 \rho}{1-2\rho} \right)\right].
\end{equation}
At low densities, this expression approaches Yang's bound for 
the condensate fraction of a generic fermionic superfluid \cite{Yang1962},
$ n_c \leq 1 - \rho$.

\begin{figure}[h!]
\includegraphics[width=12cm]{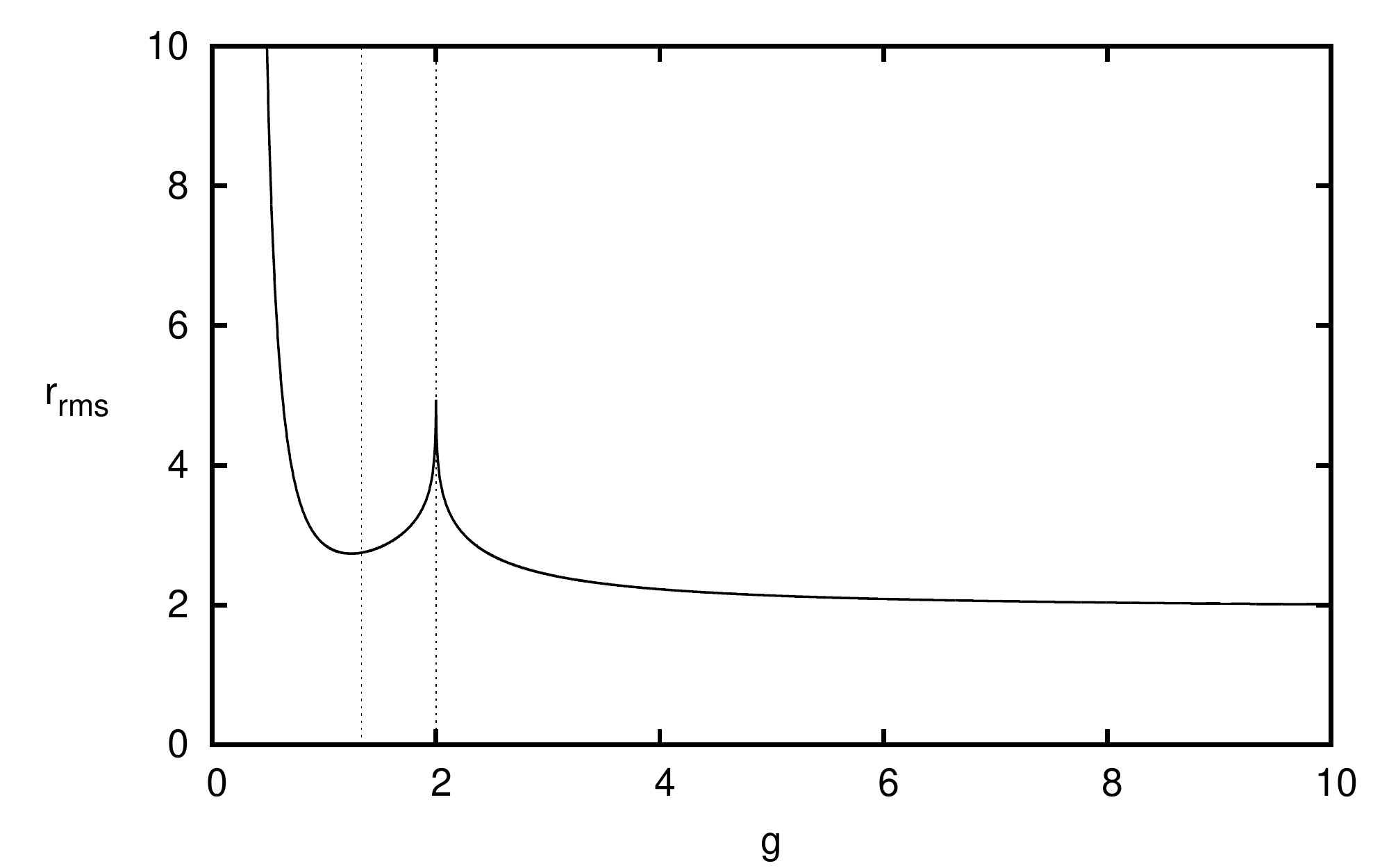}
\caption{\label{fig:radius} Root mean square radius, $r_{{\sf rms}}$,
of the condensate wave function in units of $1/\sqrt{2\omega}$,
at quarter filling. The dashed lines indicate the Moore-Read point ($\varepsilon=0$),
and the phase transition point ($\mu=0$).}
\end{figure}

Given that there is a condensate,
it is worthwhile to analyze the corresponding condensate wave function 
in momentum space,
which in the thermodynamic limit is given by \cite{Yang1962}
\begin{equation}
  \phi_{\sf Yang}(k) = u_k v_k  \propto \frac{k_x -i k_y}{\sqrt{(k^2-a)(k^2-b)}}.
\end{equation}
The condensate wave function has an inherent length scale $r_{\sf rms}$ 
that can be calculated as \cite{Ortiz2005a}
\begin{equation}
  r_{\sf rms}^2 =  \int |\nabla \phi(k) |^2 dk  /  \int |\phi(k) |^2 dk.
\end{equation}
After some algebra, one obtains the analytical expression:
\begin{equation}
  r_{\sf rms,\, Yang}^2 =  \frac{
     2\omega \left[ ab-(a+b) \omega\right] \left[  \frac{1}{(2\omega-a)^2} +  \frac{1}{(2\omega-b)^2} \right]
    + \frac{a^2-4ab+b^2}{a-b}\left[ \ln\left( 1-{2\omega}/{a}\right) - \ln\left( 1-{2\omega}/{b}\right) \right] }
  {(a-b)\left[ a \ln\left(1-{2\omega}/{a}\right) - b \ln\left(1-{2\omega}/{b}\right) \right] }.
\end{equation}
The resulting radius $r_{\sf rms,\, Yang}$ is plotted in Fig.~\ref{fig:radius}.
One sees that it goes to a finite value in the strong coupling limit,
and that it diverges at the QPT,
with a singularity in the function $\ln(1-2\omega/a)$, that scales as $\ln|\mu|$.
The condensate wave function can be related directly to observable quantities:
from Wick's theorem for the mean field solution one obtains
\begin{equation}
 \left\langle  c^\dagger_{r} c^\dagger_{r'} \right\rangle
 \left\langle  c^{\phantom{\dagger}}_{r'} c^{\phantom{\dagger}}_{r} \right\rangle 
 =   \left\langle  c^\dagger_{r}  c^\dagger_{r'}
       c^{\phantom{\dagger}}_{r'} c^{\phantom{\dagger}}_{r}  \right\rangle
   - \left\langle  c^\dagger_{r}  c^{\phantom{\dagger}}_{r\phantom{'}}  \right\rangle
     \left\langle  c^\dagger_{r'} c^{\phantom{\dagger}}_{r'} \right\rangle
   + \left\langle  c^\dagger_{r}  c^{\phantom{\dagger}}_{r'} \right\rangle
     \left\langle  c^\dagger_{r'} c^{\phantom{\dagger}}_{r}  \right\rangle.
\end{equation}
We can rewrite this as
\begin{equation}
 \left| \phi_{\sf Yang}(r-r') \right|^2
 \propto \left\langle \left( n_r - \langle n_r \rangle \right) ^{\phantom \dagger}
                \left( n_{r'} - \langle n_{r'} \rangle \right) \right\rangle
   +  \left|  F(r-r') \right|^2,
   \label{eq:phyr}
\end{equation}
where $\phi_{\sf Yang}(r)$ is the condensate wave function in coordinate space
and $F(r)$ is the Fourier transform of the momentum density
$\left\langle c^\dagger_k c^{\,}_k\right\rangle $.
In trapped Fermi gases the first term of the right hand side of Eq.(\ref{eq:phyr})
can be obtained through quantum noise interferometry\cite{Rom2006},
while the momentum density and hence $F(r)$ can be obtained from a time-of-flight analysis
after releasing the trap.
Once these quantities are known, one can calculate the radius $r_{\sf rms,\, Yang}$.
Therefore this length scale is an experimentally accessible quantity
that would give a clear signature
of the third-order QPT in the $p_x+ip_y$ superfluid.

We can apply a similar analysis to the Bethe ansatz solution:
in Eq.(\ref{eq:bethe_ansatz}) we see that the exact eigenstates
are given by a product of pair wave functions,
\begin{equation}
 {\sf S}^+(E_\alpha) = \sum_k \phi_{E_\alpha}(k) c_k^\dagger c_{-k}^\dagger,
    \ \ \ \mbox{with} \
    \phi_{E}(k) \propto \frac{k_x +i k_y}{k^2 - E}.
\end{equation}
The ground state  is given by a set of pairon values $E_\alpha$
and hence can be interpreted as an ensemble of Cooper pairs of varying size \cite{Ortiz2005a,Dukelsky2006}.
Evaluating the root mean square radius for the Cooper pair wave function $\phi_{E}(r)$
in coordinate space, one finds that
\begin{eqnarray}
 r_{{\sf rms},\, E}^2 & =&
  8 \frac{ 2\omega \mbox{Re}\left[E\right] ( \mbox{Re}\left[E\right] - 2\omega )
            \left( \frac{1}{2\omega - E} - \frac{1}{2\omega - E^*} \right)
         + |E|^2 \left[ \ln\left(1-{2\omega}/{E^*}\right) - \ln\left(1-{2\omega}/{E}\right) \right] }
         {(E-E^*)^2
        \left[E \ln\left(1-{2\omega}/{E}\right) - E^* \ln\left(1-{2\omega}/{E^*}\right) \right]},
  \\ & = &
      \frac{4\omega \left( 3 E^2 - 6 \omega E + 8 \omega^2\right) }
           {3 E (E-2\omega)^3 \left[ \ln(1-2\omega/E) +2\omega/(E-2\omega)\right]  },
     \ \ \ \mbox{for}\ E\ \mbox{real and negative}.
\end{eqnarray}
This length scale is finite if $E$ is complex,
and converges to a finite value when $E$ approaches the real negative axis.
But when approaching real and positive values of $E$, inside the interval $[0,2\omega]$,
the length scale $r_{{\sf rms},\, E}$ diverges.
The exact solution of the integrable $p_x+ip_y$ model gives us a beautiful insight
into the nature of the QPT:
in strong pairing all pairon values $E_\alpha$ are real and negative,
which means that the exact ground state corresponds to an ensemble
of confined Cooper pairs with finite radius.
Lowering the interaction strength one encounters the QPT
at which one pairon value becomes real and positive,
which corresponds to a single deconfined Cooper pair
on top of the ensemble of confined Cooper pairs.
Entering further into the weak-pairing phase,
the pairon values become complex.
At integer values of $Q$ there are $2Q+1$ deconfined Cooper pairs,
while at intermediate values
part of the pairons form two arcs around the origin in the complex plane,
some stick to the lowest part of the real positive axis (inside the arcs)
and the remaining ones stay confined on the real negative axis.
At the Moore-Read line all Cooper pairs are deconfined.
From there on part of the Cooper pairs will form resonances
of finite size (complex pairon values),
till all Cooper pairs convert to plane waves in the extreme weakly interacting limit.
Hence we can understand the QPT as a confinement-deconfinement phase transition.

%==============================================================================
\section{Symmetries of the Richardson-Gaudin equations}
%==============================================================================
\label{sec:symmetry}

In this section we highlight some symmetry relations of the RG equations
that allow us to relate states in the strong-pairing phase
with mirror states in the weak-pairing range that have the same energy\cite{Ibanez2009}.
To set the stage we would like to introduce a new set of operators,
\begin{eqnarray}
\{T^+_m , T^-_m, T^z_m \} , \ \ \mbox{with}\ m = \cdots, -2,-1,0,1,2, \cdots ,
\end{eqnarray}
which form an affine Lie algebra
\begin{equation}
\left[ T^+_m,T^-_{m^{\prime}} \right] = 2T^z_{m+m^{\prime}}, \ \ \ \left[
T^z_m , T^{\pm}_{m^{\prime}} \right] = \pm T^{\pm}_{m+m^{\prime}}.
\label{affine_lie}
\end{equation}
A possible representation of these elements of the algebra in terms of
$su(2)$ spin operators is ($\sigma=z,\pm$)
\begin{equation}
T^\sigma_m = \sum_i (\eta_i)^{m/2} S^\sigma_i,
\end{equation}
for (in principle arbitrary) quantum number $i$.

It is straightforward to check that the following two families of Hamiltonians
\begin{eqnarray}
  H_r &=& T^z_{2m} -G  \ T^+_0 T^-_0 ,
 \\
  H_h &=& T^z_{2m} -G  \ T^+_m T^-_m ,
\end{eqnarray}
define the integrable rational and hyperbolic RG models of
Eq.(\ref{eq:sp_hamiltonian}).
It is illuminating to solve those integrable rational and hyperbolic models in
terms of the generators of the affine Lie algebra. The eigenvectors can
generically be written as in Eq. (\ref{eq:bethe_ansatz}) with (unnormalized)
\begin{eqnarray}
 \mathsf{S}^{\pm}(E_{\alpha })
  &=& \sum_{m=0}^{\infty } \frac{G}{E_{\alpha }^{m+1}} \ T_{2m}^{\pm}\ ,
      \ E_{\alpha }\neq 0\ ,\ \mathsf{S}^{\pm}(0)=T_{-2}^{\pm}\ ,\ \mbox{ rational model},
 \\
 \mathsf{S}^{\pm}(E_{\alpha })
  &=& \sum_{m=0}^{\infty } \frac{2\gamma\sqrt{E_{\alpha}}}{E_{\alpha }^{m+1}}\ T_{2m+1}^{\pm}\ ,
      \ E_{\alpha }\neq 0\ ,\ \mathsf{S}^{\pm}(0)=T_{-1}^{\pm}\ ,\
\mbox{ hyperbolic model} ,
\end{eqnarray}
where $E_\alpha$ are the pairons. Notice that to represent these eigenvectors one
needs an infinite set of generators, the even series for the rational model and the odd
for the hyperbolic model.
In terms of the affine operators one can demonstrate some interesting symmetry
properties of the model:

{\sf  Theorem 1}: Given the hyperbolic Hamiltonian $H_h$, and a particular eigenstate
$\left\vert \Phi_{M} \right\rangle$ of order $M \leq (L-G^{-1})/2$, for
integer values of $G^{-1}=L-2M+1-N$, there exists another eigenstate $
\left\vert \Phi_{M+N} \right\rangle  = \left(T_{-1}^{+}\right)^{N}
\left\vert \Phi_{M} \right\rangle$ with the same energy $E$.

{\sf  Proof}: Define $H_h=T^z_2- G \ T^+_1 T^-_1$. One can straightforwardly  evaluate the
commutators
\begin{eqnarray}
\left[ H_h,T_{-1}^{+} \right] &=& T_{1}^{+}+2G T_{1}^{+}T_{0}^{z}, \\
\left[ \left[ H_h,T_{-1}^{+}\right], T_{-1}^{+} \right] &=& 2G
T_{1}^{+}T_{-1}^{+}.
\end{eqnarray}
Now let us consider a state of the form
\begin{eqnarray}
\left\vert \Phi_{M+N} \right\rangle  = \left( T_{-1}^{+} \right)^{N}
\left\vert \Phi_{M} \right\rangle.
\end{eqnarray}
We want to determine the conditions for $\left\vert \Phi_{M+N} \right\rangle$ to be
an eigenstate if $\left\vert \Phi_{M} \right\rangle$ is another
eigenstate of $H_h$ of energy $E$. Using the commutation relations above we have
\begin{eqnarray}
\left[ H_h,\left( T_{-1}^{+}\right)^{N} \right]
   = N \left( T_{-1}^{+} \right)^{N-1} T_{1}^{+}  \left[ 1+2G T_0^z + G(N - 1) \right].
\end{eqnarray}
Therefore
\begin{eqnarray}
H_h \left\vert \Phi_{M+N} \right\rangle  = E
\left\vert \Phi_{M+N} \right\rangle  + N \left( T_{-1}^{+} \right)^{N-1}
T_{1}^{+} \left[ 1  + G (2M - L + N-1) \right] \left\vert
\Phi_M\right\rangle.
\end{eqnarray}
The condition to be an eigenstate is
\begin{equation}
G^{-1} = L-N-2M+1.
\end{equation}
This means that for a certain value of $G=1/(L-M-M^{\prime}+1)$, there exist
two states, one of order $M$ and another state of order $M^{\prime}$,
that have the same energy $E$.
{} From this algebraic relation, one can distinguish 
the two special cases already mentioned in Sec.~\ref{sec:singularity}:
\begin{description}
\item[Special case (i)]  A first case is the symmetric case, that occurs
when $M=M^{\prime}$,  i.e. when
\begin{equation}
M=(L+1-G^{-1})/2, \ \ \ Q= -\frac{1}{2};
\end{equation}
\item[Special case (ii)]  Another special case occurs when $M^{\prime}=0$,
which amounts to
\begin{equation}
M=L+1-G^{-1}, \ \ \ Q= \frac{M-1}{2}.
\end{equation}
The corresponding eigenstate of order $M$ is given by
 $ \left\vert 0_M \right\rangle  = \left( T_{-1}^{+} \right)^{M} \left\vert \nu \right\rangle$.
The energy of this state is equal to the vacuum energy,
$E(0_M)=\langle \nu |H_h| \nu \rangle$.
Comparing this state with the Bethe ansatz from Eq.(\ref{eq:bethe_ansatz}),
one observes that this state corresponds to the case  where the $M$ pairons converge to $0$.
\end{description}

These algebraic relations are easily recognized to be symmetries of the RG equations as well.
Consider Eqs. (\ref{eq:sp_eigeneq}) and (\ref{eq:q}) for $G^{-1}=L-2M+1-N$,
and evaluate (\ref{eq:sp_eigeneq}) and (\ref{eq:q}) for the cases with $M$  and
$M+N$ pairs, where $N$ pairons $E_\beta$, $\beta=M+1,\cdots, M+N$,  are zero.
Then, $Q$ for the $M$ pair case is given by $Q=-(N+1)/2$ while for the $M+N$ situation it is
$Q=-(N+1)/2+N$. On the other hand, the first term in Eq. (\ref{eq:sp_eigeneq}) is the
same for $M$ or $M+N$ but the second one differs precisely by $N/E_\alpha$, so that
it cancels with the $-N/E_\alpha$ coming from the third term.
The bottom line is that the RG equations remain
the same for the two cases $M$ and $M+N$.

In the thermodynamic limit this symmetry manifests itself as a particular
transformation of the chemical potential $\mu$ and gap $\Delta$ which
preserves the form of the Eqs.(\ref{eq:analytic_meanfield1}) and
(\ref{eq:analytic_meanfield2}). For certain values of $\mu$ and
$\Delta^2$ there exists another set of values $\mu^{\prime}$ and $
\Delta^{\prime 2}$ that lead to the same equations,
\begin{eqnarray}
\mu^{\prime}& = & - \mu, \\
\Delta^{\prime 2} & = & \Delta^2 - 2 \mu.
\end{eqnarray}

The theorem above relates eigenstates belonging to different Hilbert subspaces for
particular values of $G$ and is a result also obtained in Ref.~\onlinecite{Dunning2010} 
in the context of the $p_x+ip_y$ superfluid.
By using the affine Lie algebra introduced in Eqs. (\ref{affine_lie}), we have generalized 
this result to arbitrary realizations and representations of the general hyperbolic RG model. 
One may wonder whether  a similar symmetry property holds for the rational RG model
$H_r=T_{2}^{z}-G \ T_{0}^{+}T_{0}^{-}$.  Let us assume that there exists a state
\begin{equation}
 \left\vert \Phi _{M+N}\right\rangle
   = \left( T_{-2}^{+} \right)^{N} \left\vert \Phi_{M} \right\rangle ,
\label{ratio_state}
\end{equation}
and determine the conditions to be an eigenstate of $H_r$
whenever $\left\vert \Phi_{M}\right\rangle$ is.
The commutator
\begin{eqnarray}
 \left[ H_r, (T_{-2}^{+})^{N} \right]
      = N \left( T_{-2}^{+} \right)^{N-1} \left[ H_r, T_{-2}^{+} \right]
              + \frac{ N \left( N-1 \right) }{2} \left( T_{-2}^{+} \right)^{N-2} \left[ \left[H_r,T_{-2}^{+}\right], T_{-2}^{+} \right]
\end{eqnarray}
implies that the action of the Hamiltonian on the state (\ref{ratio_state}) is
\begin{eqnarray}
 H_r \left\vert \Phi_{M+N} \right\rangle
  = E \left\vert \Phi_{M+N}\right\rangle
       +N \left( T_{-2}^{+} \right)^{N-2} T_{0}^{+} \left[ T_{-2}^{+}\left(1+2G T_{-2}^{z}\right) + G \left( N-1 \right) T_{-4}^{+} \right]
        \left\vert \Phi_{M}\right\rangle
\end{eqnarray}

To cancel out the second term two conditions need to be satisfied. First, the
number of pairons with $E_\alpha=0$ must be $N=1$. Second,
\begin{eqnarray}
\left (1+2G T_{-2}^{z}\right)  \left\vert \Phi_{M}\right\rangle = 0,
\end{eqnarray}
which indicates that for a particular value of $G$, satisfying the relation
above, a state with at most {\it one} zero pairon may exist in the rational model,
as opposed to the hyperbolic  case where states with an arbitrary number of
zero pairons are possible eigenvectors. This qualitative difference reflects
in the quantum phase diagram. While the hyperbolic model displays a
QPT from a weak-pairing phase to a strong-pairing
phase, the rational model displays a crossover.  This is so, since the symmetry
relation indicates that at the $\mu=0$ line the ground state becomes degenerate,
while this is not the case in the rational model.
Connecting points with equal energies in the phase diagram,
we see that they form two lines (one in weak pairing and one in strong pairing)
that approach the phase transition line at the same point but under different angles.
Hence it is no surprise that the energy is continuous but
has a discontinuity in the third and higher derivatives at this point.

Another interesting symmetry relation, not discussed before, emerges by considering the
integrable  Hamiltonian
\begin{eqnarray}
\tilde{H}_h=T_{-2}^{z} - \tilde{G} \ T_{-1}^{+}T_{-1}^{-} ,
\end{eqnarray}
which is  another exactly solvable hyperbolic RG Hamiltonian,
corresponding to parameters $\tilde{\eta}_i=1/\eta_i$. A simple
way to derive it is to make the choice $\epsilon_i=\tilde{\lambda}\eta_i^{-1}$  in
Eq.(\ref{eq:rg_hamiltonian}).

{\sf  Theorem 2}: Given the hyperbolic Hamiltonian $H_h$, there exists
another $\tilde{H}_h$ with $\tilde{G}^{-1} = L - 2M + 2 -{G}^{-1}$ whose
eigenstates are the same but the eigenvalues are given by
$\tilde{E} =\langle \nu \vert \tilde{H}_h \vert \nu \rangle + \sum_\alpha E_\alpha^{-1}$.
Hence, although $\tilde{H}_h$ must have the same eigenstates as
$H_h$ for the same Hilbert subspace of order $M$, their eigenvalues are
different.

{\sf  Proof}:
We  explore now how the eigenvalues of $\tilde{H}_h$ are related to those of $H_h$.  The
commutator with $H_h$ is given by
\begin{eqnarray}
  \left[ H_h,\tilde{H}_h \right]
     = \left( T_{1}^{+} T_{-1}^{-} - T_{-1}^{+} T_{1}^{-} \right)
       \left[ 2G \tilde{G}\left( 1-T_{0}^{z}\right) - G - \tilde{G} \right].
\end{eqnarray}
Acting on a state of order $M$  it results
\begin{eqnarray}
 \left[  H_h,\tilde{H}_h \right]  \left\vert \Phi_M \right\rangle
   = \left[  G \tilde{G} \left( L - 2M +2 \right) - G - \tilde{G} \right]
    \left( T_{1}^{+}T_{-1}^{-} - T_{-1}^{+}T_{1}^{-}\right)  \left\vert \Phi_M \right\rangle.
\end{eqnarray}
Therefore, for given values of $G$ and $M$ the relation
\begin{eqnarray}
\tilde{G}^{-1} +{G}^{-1}  = L - 2M + 2,
\end{eqnarray}
defines a value of $\tilde{G}$ such that $[ H_h,\tilde{H}_h] =0$.  This
algebraic identity also shows up in Eq.(\ref{eq:sp_eigeneq}),  since these equations
are invariant under the transformation
\begin{equation}
\tilde{\eta}_i = \eta_i^{-1} , \ \ \ \tilde{E}_\alpha = E_\alpha^{-1} , \ \ \
\tilde{G}^{-1} = L - 2M + 2 -{G}^{-1} .
\end{equation}
Hence one can conclude that for a given eigenstate
$\left\vert \Phi_M \right\rangle$ of $H_h$,
the following relation holds:
\begin{equation}
\langle \Phi_M \vert \tilde{H}_h \vert \Phi_M \rangle =
\langle \nu \vert \tilde{H}_h \vert \nu \rangle + \sum_\alpha
E_\alpha^{-1}.
\end{equation}
In particular, at $G^{-1}=L-2M+2$,
the state $\left\vert \Phi_M \right\rangle$
becomes an eigenstate of the operator $T_{-1}^{+}T_{-1}^{-}$.

%==============================================================================
\section{Conclusions}
%==============================================================================
\label{sec:conclusions}

The exactly-solvable hyperbolic RG model
allows a parameterization that leads to a pairing model
for spinless fermions in two dimensions
with a $p$-wave interaction of $p_x +i p_y$ character.
The model can be solved with a Bethe ansatz,
even for large system sizes,
using a new numerical technique presented here.
The thermodynamic limit can be evaluated analytically,
leading to expressions for the ground state energy and few-body correlation functions 
that coincide with mean field theory.

Even though the $p_x +i p_y$ model has been studied before~\cite{Ibanez2009,Dunning2010},
we reach drastically different conclusions about the zero temperature quantum phase diagram.
Our numerical and analytical results
show that the quantum phase diagram exhibits
a third-order quantum phase transition when the chemical potential changes sign.
Most importantly, as shown in this paper, this transition is characterized by a 
logarithmically diverging length scale
that can be obtained experimentally from correlations in the density fluctuations.
This length scale can be associated with the size of the Cooper pairs making up the correlated many-body state.
Those Cooper pairs, tightly bound in the strong-pairing phase,
start to  deconfine at the QPT point where $\mu=0$.
Thus, in a certain sense, our physical picture is that the strong-pairing region represents a confined superfluid phase
while the weak-pairing region is the deconfined phase.
Similarly to many other examples of phase transitions which are not of the Landau type,
as for instance in lattice gauge theories,
the transition is not associated with any broken symmetry and thus has no local order parameter.

We show that these are general features of the hyperbolic RG model,
irrespective of the specific symmetry or dimensionality of the representation.
This means that our conclusions equally apply
to other models derived from the same generalized Gaudin algebra,
be it models for pairing in atomic nuclei,
or Jaynes-Cummings like models of resonantly coupled fermionic atoms.
Furthermore, the algebraic underpinnings of the hyperbolic RG model
allow us to demonstrate that there exists a transformation that relates states
in strong pairing with states in weak pairing,
and that the phase transition line corresponds
to the states that are symmetric under this transformation.

Finally, we would like to emphasize the fact
that the phase diagram is based on the exact solution,
which guarantees that the phase transition is not an artefact of the mean field approximation.
Moreover, the Hamiltonian (\ref{eq:hamhyperbolic}) is integrable for attractive and repulsive interactions
while there is no mean field solution in the yet unexplored repulsive case.
Therefore, exact solvability could be a unique tool to investigate repulsive $p$-wave interactions.

%------------------------------------------------------------------------------
\begin{acknowledgments}
We thank M. Iba\~{n}ez, G. Sierra and R. Molina for  interesting discussions.
This work was supported by the Lundbeck Foundation,
a Marie Curie Intra European Fellowship (PIEF-GA-2008-220335/PHY) from the European Commission
and a Project for Fundamental Research (FIS2009-07277) from the Spanish Ministry for Science and Innovation.

\end{acknowledgments}
%------------------------------------------------------------------------------

\appendix

%==============================================================================
\section{One pair}
%==============================================================================
\label{app:onepair}

Already the state with one pair leads to a non-trivial result.
Equation (\ref{eq:sp_eigeneq}) reduces to:
\begin{equation}
   \frac{Q}{E_1} + \frac{1}{2}\sum_{k, k_x>0}  \frac{1}{E_1 - \eta_k}
          =0.
  \label{eq:sp_eigeneq1}
\end{equation}
In order to solve this equation,
we have to specify the space in which the model is defined.
Let us consider a disk in momentum space,
with radius $aR$ such that $R$ is integer and $a$ is the spacing
between successive momentum states.
The kinetic energy dispersion is given by $k^2/2$,
which means that there is a kinetic energy cutoff $\omega = (aR)^2/2$.
In the continuum limit, corresponding to $R \gg 1$,
one can take $L=\pi R^2/2$
and replace the sums over $k$ by integrals with $dk_x=dk_y=a$:
\begin{eqnarray}
  \sum_{k, k_x>0} f(k^2) & \cong &
      \frac{1}{a^2} \iint_{k_x>0,k<aR} f(k^2) dk_x dk_y
  \nonumber \\
     & = &   \frac{\pi}{a^2} \int_{k=0}^{aR} f(k^2) kdk
  \nonumber \\
     & = &   \frac{\pi R^2/2}{(aR)^2} \int_{0}^{(aR)^2} f(\eta) d\eta
  \nonumber \\
     & = &   \frac{L}{2\omega} \int_{0}^{2\omega} f(\eta) d\eta.
  \label{eq:thermodynamiclimit}
\end{eqnarray}
Hence,
\begin{eqnarray}
  \sum_{k, k_x>0}  \frac{1}{E_1 - k^2}
     & = & \frac{L}{2\omega} \int_{0}^{2\omega} \frac{1}{E_1 - \eta} d\eta
  \nonumber \\
     & = & -\frac{L}{2\omega} \ln \left( 1 - \frac{2 \omega}{E_1} \right).
\end{eqnarray}
Substituting this result in Eq.(\ref{eq:sp_eigeneq1}), one obtains that
\begin{equation}
   2(Q/L) \frac{2\omega} {E_1} - \ln \left( 1 - \frac{2 \omega}{E_1} \right)
          =0.
  \label{eq:richgaudin1solve}
\end{equation}
In the limit $L \gg 1$ one finds that $E_1= 2\omega /f_g$,
where $g=GL$ and $f_g$ is the solution of the following equation:
\begin{equation}
   (1/g-1) f_g = \ln \left( 1 - f_g \right).
\end{equation}

If $1/g < 1$, then $f_g < 0$
and one obtains a ground state energy for one pair,
\begin{equation}
 E_1 = 2\omega /f_g,
\end{equation}
such that $E_1 < 0$.
This value $E_1$ then defines a physical energy scale
that can be used to eliminate the cutoff $\omega$ as done in Ref.~\onlinecite{Botelho2005}.
For $1/g \geq 1$, the one-pair ground state corresponds
to a non-paired state with $E_1=0$.
In this case there is no relevant energy scale to renormalize the cutoff.

The one-pair energy is a physically meaningful quantity,
that allows one to relate different models describing the same physics.
Typically, models with $p_x+ip_y$ pairing will have an interaction
that scales as $(k_x+ik_y)(k'_x-ik'_y)$ for small $k$,
just like the integrable $p_x+ip_y$ model,
but for large momenta the interaction could be very different.
Because the peculiarities of the quantum phase diagram
are determined mainly by the small-$k$ behavior,
we expect similar results for other two-dimensional $p_x+ip_y$ models,
as e.g. in Ref.~\onlinecite{Botelho2005}.
To relate their results to ours, one has to renormalize the interaction strength,
which can be done by equating the one pair energy.

%==============================================================================
\section{Polynomial solution for the one-level model}
\label{app:onelevel}
%==============================================================================

In the limiting case of a single level, $l=1$,  Eq.(\ref{eq:sp_eigeneq}) takes the form:
\begin{equation}
  \frac{s}{\eta - E_\alpha}
      - \sum_{\alpha', \alpha' \neq \alpha}\frac{1}{E_{\alpha'}-E_{\alpha}}
         = \frac{Q}{E_\alpha},
         \ \ \forall \alpha.
 \label{eq:1l_eigeneq}
\end{equation}
Equation (\ref{eq:1l_eigeneq}) is a generalized Stieltjes equation \cite{Shastry2001},
that can be solved by multiplying it
by  $2E_\alpha \left(\eta - E_\alpha\right) P'_M(E_\alpha)$,
where $P_M(x)=\prod_{\alpha'=1}^M (x-E_{\alpha'})$
is the polynomial that has the unknowns $E_\alpha$ as its roots.
One finds that
\begin{equation}
   \left[ 2(s+Q) E_\alpha - 2Q \eta \right] P'_M(E_\alpha)
     + E_\alpha \left(\eta - E_\alpha\right) P''_M(E_\alpha) = 0,
\end{equation}
for $\alpha=1, \ldots, M$.
Since this is a polynomial expression of order $M$,
valid in $M$ distinct points (assuming that none of the $E_\alpha$ coincide),
one can identify the polynomial expression with a multiple of $P(x)$:
\begin{equation}
   \left[ 2(s+Q) x - 2Q \eta \right] P'_M(x) + x \left(\eta - x\right) P''_M(x)
     = M\left[2(s+Q)-M+1\right]  P_M(x), \ \ \ \forall x.
 \label{eq:1l_diffeq}
\end{equation}
This equation can be transformed into the Jacobi equation in the variable $y$
where $x=(\eta/2)(1+y)$.
The solution is a Jacobi polynomial:
\begin{equation}
  P_M(x) \propto {\cal P}^{\alpha,\beta}_M (2x/\eta-1),
\end{equation}
with parameters $\alpha=2Q/\eta-2Q-2s-1$, $\beta=-2Q/\eta-1$.

Let us expand $P_M(x)=\sum_{i=0}^M p_{M,i} \, x^i$, where $p_{M,M}=1$.
By equating orders of $x^l$ in Eq.(\ref{eq:1l_diffeq}),
one obtains a recurrence relation that defines the coefficients $p_{M,i}$ completely:
\begin{equation}
  p_{M,i-1} = \eta \frac{(2Q+1-i)i}{\left[M+i-2-2(s+Q)\right](M+1-i)} p_{M,i}, \ \ \ \mbox{for}\ i=M, M-1, \ldots, 1.
  \label{eq:1l_recurrencepk}
\end{equation}
If $2Q+1$ equals an integer value $N\leq M$,
one sees that $P_M(x)$ has a factor $x^N$.
This means that a singularity occurs,
in which $N$ pairons will collapse at $0$.
Of particular interest are the cases $N=M$, were all pairons collapse to zero,
and $N=0$, where no more singularities occur.

For multilevel systems, e.g. a system of $l$ levels,
the polynomial $P_M(x)$ is still well defined.
Rather than a Jacobi-like equation, it now has to fulfill a
{\em generalized Lam\'{e} differential equation} with negative residues \cite{Marcellan2007},
\begin{equation}
  P''_M(x) + \left( \sum_{k=0}^l \frac{r_k}{x-\eta_k} \right) P'_M(x)
    - \frac{V_M(x)}{\prod_{k=0}^l \left( x- \eta_k \right) } P_M(x) =0,
\end{equation}
with $r_0=-2Q, r_k=-2s_k,$ and $ \eta_0=0 $.
The solutions $P_M(x)$ are known as {\em Stieltjes polynomials},
while the polynomials $V_M(x)$ of order $l-1$ or less
are known as {\em Van Vleck polynomials}.
If all $r_k$ were positive, i.e. all $s_k$ and $Q$ negative,
which can occur for $su(1,1)$ representations (e.g. boson pairs),
then the polynomials $P_M(x)$ would be orthogonal polynomials with respect to some weight function.
The case of negative residues, $r_k<0$,
is much less understood and a topic of ongoing research \cite{Marcellan2007}.

%==============================================================================
\section{Continuum limit of the integrable $p_x+ip_y$ model}
%==============================================================================
\label{app:continuum}

A full analysis of the continuum limit of the integrable $p_x+ip_y$
has been presented in Refs.~\onlinecite{Amico2002,Dunning2010}.
We give here a brief overview of the procedure,
that applies to the hyperbolic RG model in general,
in order to maintain consistency with the other sections
and to demonstrate that the transitions
between the different regions in Fig. \ref{fig:hyperbolic_phasediagram}
are genuine phase transitions and not merely crossovers.
The continuum limit of the RG equations (Eq.(\ref{eq:sp_eigeneq})),
is based on the following scalings:
\begin{eqnarray}
   L, M & \rightarrow & \infty,
 \\
   M/L & = & \rho \ \ \ \mbox{finite},
 \\
   GL & = & g \ \ \ \mbox{finite},
 \\
   Q/L & = & q = \frac{1}{2g} - \frac{1}{2} + \rho,
 \\
   \frac{2}{L} \sum_k s_k f(\eta_k)
     & \equiv & \int_\Lambda \varrho(\eta) f(\eta) d\eta,
\\
   \frac{1}{L} \sum_\alpha f(E_\alpha)
     & \equiv & \int_\Omega p(x) f(x) dx,
\end{eqnarray}
where $\Lambda$ is the support of the weighted level density $\varrho(\eta)$
and $\Omega$ the support of the pairon density $p(x)$.
Their normalization is given by
\begin{eqnarray}
   \int_\Lambda \varrho(\eta) d\eta &=& 1,
 \\
   \int_\Omega p(x) dx &=& \rho.
\end{eqnarray}
The RG equations Eq.(\ref{eq:sp_eigeneq})
take the following form in the continuum limit:
\begin{equation}
    \frac{q}{x} + \frac{1}{2} \int_\Lambda \frac{\varrho(\eta)}{x-\eta} d\eta
    - {\cal P} \int_\Omega \frac{p(x')}{x-x'} dx' = 0,
    \ \ \forall x \in \Omega.
 \label{eq:cont_eigeneq}
\end{equation}
The integral over $x'$ is to be understood as a Cauchy principal value, ${\cal P}$.
To solve this equation, we start from an electrostatic analogy
that goes back to work of Stieltjes on orthogonal polynomials \cite{Dimitrov2000,Marcellan2007},
and has been applied in the present context by Gaudin \cite{Gaudin1995}
and was later elaborated in Refs.~\onlinecite{Dukelsky2002,Roman2002} for the rational model
and in Refs.~\onlinecite{Amico2002,Dunning2010} for the hyperbolic model.
One considers the pairons to represents $M$ free charges
with unit negative charge, positioned in the two-dimensional plane
at coordinates given by the real and imaginary parts of pairon values $E_\alpha$.
Then Eq. (\ref{eq:cont_eigeneq}) represents the two dimensional electrostatic force generated by the repulsion among the free charges,
the attraction of a fixed charge density $\varrho(\eta)$ along the interval $\Lambda$,
and the interaction with a central charge $q$ located at the origin,
derived from a potential $V(z)$ in the two-dimensional complex plane,
\begin{equation}
 V(z)= - q \ln(|z|)
       - \frac{1}{2} \int_\Lambda \varrho(\eta) \ln(|z-\eta|) d\eta
       + \int_\Omega p(x') \ln(|z-x'|) dx'.
\end{equation}
Therefore, any particular solution of (\ref{eq:cont_eigeneq})
gives the equilibrium positions of the pairon density $p(x)$ in the two dimensional plane.
Note that $q$ can change from attractive to repulsive 
depending on the filling $\rho$ and on the coupling $g$.
This change in sign is associated with a QPT
whose interpretation is one of the main goals of this work.
\begin{figure}[h!]
\includegraphics{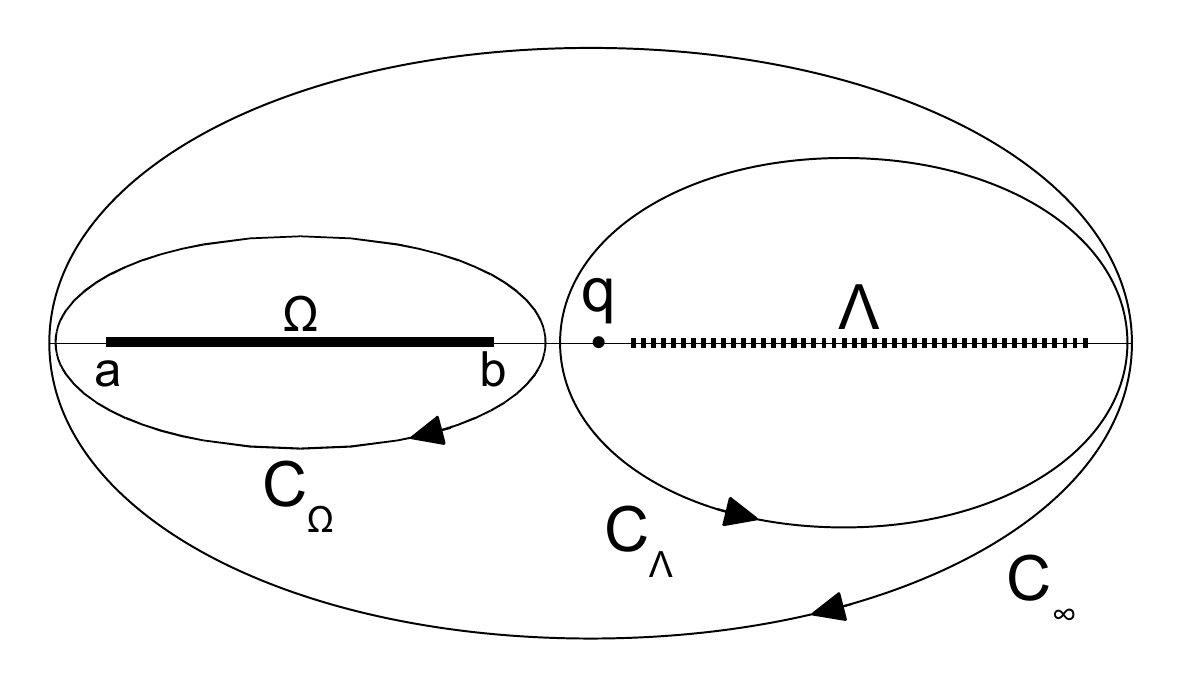}
\caption{\label{fig:contours}
 Contours used to evaluate the continuum limit.}
\end{figure}
To find the unknown pairon density $p(x)$,
we assume that it corresponds to a continuous charge distribution
along a certain curve (or set of curves) $\Omega$.
The electric field $F(z)$ created by the charge $q$
at the origin and the charge densities $p$ (repulsive) and $\varrho$ (attractive)
will have a discontinuity across $\Omega$ given by
\begin{equation}
 p(x) = \frac{F(x_+)-F(x_-)} {2 \pi i}, \ \ \ \mbox{for} \ x \in \Omega.
\end{equation}
This allows us to rewrite the Cauchy principal value integral over $\Omega$
as an integral along a clockwise contour $C_\Omega$ around $\Omega$:
\begin{equation}
  {\cal P}  \int_\Omega \frac{p(x')}{x-x'} dx'
   = \frac{1}{2\pi i} \oint_{C_\Omega} \frac{F(z)}{x-z} dz.
 \label{eq:cont_ploop1}
\end{equation}
To model the discontinuity in $F(z)$ we propose the following ansatz:
\begin{equation}
 F(z) = R_\Omega(z) \int_\Lambda \frac{\phi(\eta)}{z-\eta}d\eta,
\end{equation}
where $R_\Omega(z)$ is a function
that explicitly takes into account a cut along $\Omega$.
If $\Omega$ corresponds to a single segment with endpoints $a$ and $b$,
then one can take
\begin{equation}
 R_{a-b}(z) = \sqrt{(z-a)(z-b)}.
 \label{eq:cont_r1}
\end{equation}
For more segments, $R_\Omega(z)$ has to be a higher order function, e.g.
\begin{equation}
   R_\Omega(z) = \prod_{i=1}^n \sqrt{(z-a_i)(z-b_i)},
 \label{eq:cont_rn}
\end{equation}
where $n$ is the number of segments.
In general, one can write an asymptotic expansion for large $z$ as
\begin{equation}
   R_\Omega(z) = \sum_{i=0}^{+\infty} r_i z^{n-i}.
 \label{eq:cont_rexpansion}
\end{equation}
Note that the total charge in the system is given by $\rho-q-1/2=-1/(2g)$.
Therefore the asymptotic behavior of $F$ is given by
\begin{equation}
 F(z) = \frac{1}{2gz} + {\cal O}\left(z^{-2}\right),
   \ \ \ \mbox{for} \ |z| \rightarrow \infty.
\end{equation}
This allows us to put constraints on the unknown function $\phi(z)$:
\begin{eqnarray}
   \int_\Lambda \phi(\eta) \eta^i d\eta &=& 0, \ \ \ \mbox{for}\ i=0, \ldots, n-1;
   \label{eq:cont_cond1}
 \\
   \int_\Lambda \phi(\eta) \eta^n d\eta &=&  \frac{1}{2g}.
   \label{eq:cont_cond2}
\end{eqnarray}
Now we can rewrite Eq.(\ref{eq:cont_ploop1})
by deforming the contour around $\Omega$
into a contour $C_\Lambda$ around $\Lambda$
plus a contour $C_\infty$ around the point at infinity:
\begin{eqnarray}
  \lefteqn{ {\cal P}  \int_\Omega \frac{p(x')}{x-x'} dx'
  %\nonumber \\ & = &
         =  \frac{1}{2\pi i} \oint_{C_\Omega} \frac{R_\Omega(z)}{x-z}
            \int_\Lambda \frac{\phi(\eta)}{z-\eta}d\eta dz
  }
  \nonumber \\
    & = & \frac{1}{2\pi i} \int_\Lambda \phi(\eta)
            \oint_{C_\Lambda} \frac{R_\Omega(z)}{(z-\eta)(x-z)} dz  d\eta
  %\nonumber \\ & &
          + \frac{1}{2\pi i} \int_\Lambda \phi(\eta)
            \oint_{C_\infty}  \frac{R_\Omega(z)}{(z-\eta)(x-z)} dz d\eta
  \nonumber \\
    & = & \int_\Lambda \frac{\phi(\eta)R_\Omega(\eta)}{x-\eta} d\eta.
 \label{eq:cont_ploop2}
\end{eqnarray}
where the integral along $C_\infty$ vanishes  because of Eq.(\ref{eq:cont_cond1}) as can be seen in the following expansion:
\begin{eqnarray}
  \int_\Lambda \phi(\eta) \oint_{C_\infty}  \frac{R_\Omega(z)}{(z-\eta)(x-z)} dz d\eta
    & = & - \int_\Lambda \phi(\eta)
            \oint_{C_\infty}  \frac{\sum_{j=0}^{+\infty} r_j z^{n-j-2}}{(1-\eta/z)(1-x/z)} dz  d\eta
  \nonumber \\
    & = & - \sum_{j,k,l=0}^{+\infty} x^l r_j \int_\Lambda \phi(\eta) \eta^k d\eta
                            \oint_{C_\infty} z^{n-j-k-l-2} dz
  \nonumber \\
    & = & 2\pi i \sum_{j=0,k=0,l=0}^{+\infty} x^l r_j \delta_{k,n-1-j-l} \int_\Lambda \phi(\eta) \eta^k d\eta
  \nonumber \\
    & = & 0,
 \label{eq:cont_cinfty}
\end{eqnarray}
because only terms with $k<n$ are allowed by the Kronecker delta.
Comparing Eq.(\ref{eq:cont_ploop2}) with Eq.(\ref{eq:cont_eigeneq}),
we find that the solution is given by
\begin{equation}
 \phi(\eta)= \frac{q\delta(\eta) + \frac{1}{2}\varrho(\eta)}{R_\Omega(\eta)}.
\end{equation}
The segments of $\Omega$ correspond to equipotential curves
in the complex plane, determined by
\begin{equation}
{\rm Re} \int_{a_n}^{z_n} F(z) dz = 0.
\end{equation}
The energy density is given by $\varepsilon=E/L$ and can be evaluated as
\begin{eqnarray}
 \varepsilon &=& \langle \nu | H_h | \nu \rangle/L + \int_\Omega x p(x) dx
  \nonumber \\
%   &=&  \langle \nu | H_h | \nu \rangle/L
%        + \frac{1}{2\pi i} \oint_{C_\Omega} z F(z) dz
%  \nonumber \\
   &=&  - \frac{1}{2} \int_\Lambda \eta \varrho(\eta) d\eta
        + \frac{1}{2\pi i} \oint_{C_\Omega} z R_\Omega(z) \int_\Lambda \frac{\phi(\eta)}{z-\eta} d\eta dz
  \nonumber \\
   &=&  - \frac{1}{2} \int_\Lambda \eta \varrho(\eta) d\eta
        + \frac{1}{2\pi i} \oint_{C_\Lambda} z R_\Omega(z) \int_\Lambda \frac{\phi(\eta)}{z-\eta} d\eta dz
        + \frac{1}{2\pi i} \oint_{C_\infty} z R_\Omega(z) \int_\Lambda \frac{\phi(\eta)}{z-\eta} d\eta dz
  \nonumber \\
 %  &=&  - \frac{1}{2} \int_\Lambda \eta \varrho(\eta) d\eta
 %       + \int_\Lambda \eta R_\Omega(\eta) \phi(\eta) d\eta
 %       + \frac{1}{2\pi i}  \int_\Lambda \phi(\eta) \oint_{C_\infty} \frac{R_\Omega(z)}{1-\eta/z} dz d\eta
 % \nonumber \\
   &=&  - \frac{1}{2} \int_\Lambda \eta \varrho(\eta) d\eta
        + \frac{1}{2} \int_\Lambda \eta \varrho(\eta) d\eta
        + \frac{1}{2\pi i}  \sum_{j,k=0}^{\infty} r_j \int_\Lambda \phi(\eta) \eta^k d\eta \oint_{C_\infty} z^{n-j-k} dz
  \nonumber \\
%   &=& -  \sum_{j,k=0}^{\infty} r_j \delta_{k,n+1-j} \int_\Lambda \phi(\eta) \eta^k d\eta
%  \nonumber \\
%   &=& -  r_0 \int_\Lambda \phi(\eta) \eta^{n+1} d\eta - r_1  \int_\Lambda \phi(\eta) \eta^{n} d\eta
%  \nonumber \\
   &=& - \frac{1}{2} \int_\Lambda \varrho(\eta) \frac{\eta^{n+1}}{R_\Omega(\eta)} d\eta
       + \sum_{i=1}^n \frac{ (a_i + b_i)}{4g}.
  \label{eq:cont_energy}
\end{eqnarray}

Let us analyze in particular the case of a single segment, i.e. $n=1$.
Eqs.(\ref{eq:cont_cond1},\ref{eq:cont_cond2}) reduce to
\begin{eqnarray}
   \frac{\sqrt{ab}}{2} \int_\Lambda \frac{\varrho(\eta)}{\sqrt{(\eta-a)(\eta-b)}} d\eta
     &=& -q = \frac{1}{2} - \frac{1}{2g} - \rho,
   \label{eq:cont_meanfield1}
 \\
   \int_\Lambda \frac{\varrho(\eta)\eta}{\sqrt{(\eta-a)(\eta-b)}} d\eta
     &=&   \frac{1}{g}.
   \label{eq:cont_meanfield2}
\end{eqnarray}
It is interesting to compare this
to the result of a mean field theory for the ground state
(see Appendix \ref{app:meanfield})
based on a grand potential $H_h - \mu S^z$,
with chemical potential $\mu$ and a pairing field
$\Delta = G \sum_k \sqrt{\eta_k} \langle S^-_k \rangle$.
One obtains the same expressions
as in Eqs.(\ref{eq:cont_meanfield1},\ref{eq:cont_meanfield2}),
provided that one equates
\begin{equation}
 \sqrt{(\eta-a)(\eta-b)} = \sqrt{(\eta-2\mu)^2+4\eta\Delta^2},
\end{equation}
which means that one can identify
\begin{eqnarray}
  & a+b = 4 (\mu - \Delta^2),
    \ \ \ ab=4\mu^2 &
  \\
  & a-b = \pm 4 \Delta \sqrt{\Delta^2 - 2 \mu}, &
  \\
  & \mu = \pm \frac{\sqrt{ab}}{2},
    \ \ \
    \Delta^2=\pm \frac{\sqrt{ab}}{2} - \frac{a+b}{4} &
\end{eqnarray}
Again one can distinguish the three regions in the quantum phase diagram:
\begin{description}
 \item[weak coupling BCS:]
       $0 < \frac{\Delta^2}{2} < \mu$, and $a$ and $b$ complex.
       The pairons form an arc in the complex plane.
 \item[weak pairing at intermediate coupling:]
       $0 < \mu < \frac{\Delta^2}{2}$, and $a$ and $b$ real and negative.
       The distribution of pairons along the negative real axis between $a$ and $b$
       has to be supplemented with two arcs in the complex plane that close around the origin,
       touching the real negative axis in $a$,
       and a segment of real positive pairons between the origin
       and the point where the arcs touch the positive real axis.
 \item[strong pairing:]
       $\mu<0$, and $a$ and $b$ real and negative.
       All pairons fall in the segment on the negative real axis between $a$ and $b$.
\end{description}
For $n=1$ one can further elaborate Eq.(\ref{eq:cont_energy}) for the energy:
\begin{eqnarray}
  \varepsilon &=& \frac{\mu}{g} - \frac{\Delta^2}{g}
                  - \frac{1}{2} \int_\Lambda \varrho(\eta) \frac{\eta^2}{\sqrt{(\eta-2\mu)^2+4\eta\Delta^2}} d\eta
  \nonumber \\
%    &=& \frac{\mu}{g} - \frac{\Delta^2}{g}
%        - \frac{1}{2} \int_\Lambda \varrho(\eta) {\sqrt{(\eta-2\mu)^2+4\eta\Delta^2}} d\eta
%    \nonumber \\ & &
%        - 2(\mu-\Delta^2) \int_\Lambda \varrho(\eta) \frac{\eta}{\sqrt{(\eta-2\mu)^2+4\eta\Delta^2}} d\eta
%        + 2\mu^2 \int_\Lambda \varrho(\eta) \frac{1}{\sqrt{(\eta-2\mu)^2+4\eta\Delta^2}} d\eta,
%  \nonumber \\
%    &=& \frac{\mu}{g} - \frac{\Delta^2}{g}
%        - \frac{1}{2} \int_\Lambda \varrho(\eta) {\sqrt{(\eta-2\mu)^2+4\eta\Delta^2}} d\eta
%    \nonumber \\ & &
%        - 2(\mu-\Delta^2) \int_\Lambda \varrho(\eta) \frac{\eta}{\sqrt{(\eta-2\mu)^2+4\eta\Delta^2}} d\eta
%        + 2\mu^2 \int_\Lambda \varrho(\eta) \frac{1}{\sqrt{(\eta-2\mu)^2+4\eta\Delta^2}} d\eta,
%  \nonumber \\
%    &=& \frac{\mu}{g} - \frac{\Delta^2}{g}
%         - \frac{1}{2} \int_\Lambda \varrho(\eta) {\sqrt{(\eta-2\mu)^2+4\eta\Delta^2}} d\eta
%     \nonumber \\ & &
%        - \frac{2(\mu-\Delta^2)}{g}
%        + \mu(1/g-1+2m)
%  \nonumber \\
    &=& \mu(2\rho-1) + \frac{\Delta^2}{g} - \Theta(\mu,\Delta^2),
  \label{eq:cont1_energy}
\end{eqnarray}
where we have defined
\begin{equation}
  \Theta(\mu,\Delta^2)
     = \frac{1}{2} \int_\Lambda \varrho(\eta){\sqrt{(\eta-2\mu)^2+4\eta \Delta^2}} d\eta.
  \label{eq:cont1_theta}
\end{equation}
We can formally express Eqs.(\ref{eq:cont_meanfield1},\ref{eq:cont_meanfield2}) as
\begin{equation}
  \frac{d\Theta}{d\mu}(\mu,\Delta^2) = 2\rho-1,
  \ \ \
  \frac{d\Theta}{d\Delta^2}(\mu,\Delta^2) = \frac{1}{g}.
  \label{eq:cont1_meanfield}
\end{equation}
Substituting these conditions into Eq.(\ref{eq:cont1_energy}),
we see that they guarantee
that $\varepsilon(\mu,\Delta^2)$ reaches an extremum at the corresponding values of $\mu$ and $\Delta^2$,
which coincides with the mean field theory of Appendix \ref{app:meanfield}.

Note that the pairon distribution does not depend
on the details of the representation for the $su(2)$ operators ${\bf S}_k$.
The only model factor that enters into the equations
is the weighted level density $\varrho(\eta)$,
which corresponds to the distribution of $s_k$ in the vacuum state $|\nu\rangle$.
The fact that $\varepsilon$, $\rho$ and $g$ all depend on $\Theta(\mu,\Delta^2)$
means that a QPT will occur
only if $\Theta(\mu,\Delta^2)$ displays a singularity in one of its derivatives.
{} From Eq.(\ref{eq:cont1_theta}) one can infer that this will be the case
if the argument of the square root vanishes,
which happens at $\mu=0$ if $\eta=0$ is included in the domain $\Lambda$.

%==============================================================================
\section{Mean field theory of the hyperbolic Richardson-Gaudin model}
%==============================================================================
\label{app:meanfield}
Mean field theories for separable interactions have been discussed before
in the context of $p_x+ip_y$ pairing models \cite{Botelho2005,Ibanez2009}.
Here we present a more general derivation applicable to all exactly solvable models 
derived from the hyperbolic RG model,
the $p_x+ip_y$ model representing a particular case.
Our mean field theory for the Hamiltonian $H_h$
from Eq.(\ref{eq:sp_hamiltonian}) is based upon a variational wave function
\begin{equation}
 | \theta \rangle = e^{\sum_k \theta_k S^+_k} | \nu \rangle.
 \label{eq:mf_ansatz}
\end{equation}
For fermions, the mean field theory is often presented in terms of a double set of 
complex parameters $v_k/u_k=\theta_k$,
with the condition that $|u_k|^2+|v_k|^2=1$.
These parameters can be related to $\theta_k$ as
\begin{equation}
  u_k = \frac{1}{\sqrt{1+|\theta_k|^2}},
 \ \ \
  v_k = \frac{\theta_k}{\sqrt{1+|\theta_k|^2}},
\end{equation}
Let us denote normalized expectation values as $\left\langle \ldots \right\rangle$.
{} From the normalization condition
\begin{equation}
  \langle \theta | \theta \rangle
    = \prod_k \left(1+|\theta_k|^2 \right)^{2 s_k},
 \ \ \
  S^z_k |\nu\rangle=-s_k|\nu\rangle,
\end{equation}
one easily finds the normalized expectation values for the basic operators of the algebra:
\begin{equation}  
    \left \langle  S^z_k \right\rangle
      = -s_k  \frac{1 - |\theta_k|^2 }{1 + |\theta_k|^2 },
  \ \ \
    \left \langle  S^-_k \right\rangle
      = 2s_k  \frac{\theta_k}{1 + |\theta_k|^2 }.
\end{equation}
Furthermore, one can show that
\begin{equation}
  \left\langle S^+_k S^-_{k'} \right\rangle
     = \left\langle S^+_k \right\rangle \left\langle S^-_{k'} \right\rangle
       + \delta_{k{k'}} \frac{\left( s_k + \left\langle S^z_k \right\rangle\right) ^2}{2s_k}.
\end{equation}
This allows us to write the mean field expectation value for $H_h$ as
\begin{eqnarray}
 \left\langle H_h \right\rangle
   &=&  \sum_k \eta_k  \left\langle S^z_k \right\rangle
           -  G \sum_{k,{k'}} \sqrt{\eta_k\eta_{k'}}
              \left\langle S^+_k \right\rangle \left\langle S^-_{k'} \right\rangle
   % \nonumber \\ & &
         - G \sum_k \eta_k  \frac{\left( s_k + \left\langle S^z_k \right\rangle\right) ^2}{2s_k}.
   \label{eq:mf_energy}
\end{eqnarray}
To derive the mean field equations,
we will vary the following expression:
\begin{equation}
  \langle \theta | H_h | \theta \rangle - 2 \mu  \langle \theta | S^z | \theta \rangle - E_{\sf mf}  \langle \theta | \theta \rangle,
 \label{eq:mfvariational}
\end{equation}
where $\mu$ is a Lagrange multiplier that fixes $\left\langle S^z \right\rangle$
and represents the chemical potential in particle (fermion or boson) 
representations of the model,
while $E_{\sf mf}$ is needed to fix the normalization.
By varying Eq.(\ref{eq:mfvariational}) with respect to $\langle \theta |$ we obtain the mean field equation:
\begin{equation}
 H_{\sf mf} | \theta \rangle = E_{\sf mf} | \theta \rangle,
\end{equation}
with
\begin{eqnarray}
 H_{\sf mf}
   &=&   2 \sum_k \xi_k  S^z_k
        - \sum_k \sqrt{\eta_k} \left( \Delta S^+_k + \Delta^* S^-_k \right),
   \label{eq:mf_hamiltonian}
 \\
 \xi_k &=& \left( \frac{1}{2}-G \frac{s_k + \left\langle S^z_k \right\rangle}{2s_k}\right)  \eta_k - \mu,
  \\
 \Delta &=& G \sum_k \sqrt{\eta_k}  \left\langle S^-_k \right\rangle.
   \label{eq:mf_delta}
\end{eqnarray}
To check whether the ansatz of Eq.(\ref{eq:mf_ansatz}) is indeed an eigenstate of $H_{\sf mf}$,
we evaluate
\begin{eqnarray}
  & & H_{\sf mf} | \theta \rangle
   = e^{\sum_k \theta_k S^+_k}
        e^{-\sum_k \theta_k S^+_k}H_{\sf mf} e^{\sum_k \theta_k S^+_k} | \nu \rangle
  \nonumber \\
%    &=& e^{\sum_k \theta_k S^+_k} \left( H_{\sf mf}
%          - \sum_k \theta_k \left[S^+_k, H_{\sf mf} \right]
%          + \frac{1}{2} \sum_{k,k'} \theta_k \theta_{k'}
%          \left[S^+_k, \left[S^+_{k'},  H_{\sf mf} \right]\right]
%          \right) | \nu \rangle
%  \nonumber \\
   & &= e^{\sum_k \theta_k S^+_k} \left(
          2\sum_k \left[ \xi_k + \theta_k \sqrt{\eta_k} \Delta^*\right]S^z_k
          + \sum_k  \left[ (\theta_k^2  \Delta^* - \Delta) \sqrt{\eta_k}
          + 2\theta_k \xi_k \right] S^+_k
          - \sum_k \sqrt{\eta_k} \Delta^* S^-_k
          \right) | \nu \rangle.
\end{eqnarray}
One sees that $|\theta\rangle$ is an eigenstate provided that
\begin{equation}
  (\theta_k^2  \Delta^* - \Delta) \sqrt{\eta_k}
    +2 \theta_k \xi_k = 0.
  \label{eq:mf_eigeneq}
\end{equation}
The corresponding eigenvalue of $H_{\sf mf}$ is given by
\begin{equation}
  E_{\sf mf} = -2 \sum_k s_k \left[ \xi_k + \theta_k \sqrt{\eta_k} \Delta^* \right].
\end{equation}
$\theta_k$ must have the same phase as $\Delta$,
which we can absorb in the definition of the operator ${\bf S}_k$,
so we can take $\Delta$ and $\theta_k$ to be real
without loss of generality.
One finds that the ground state solution
of Eq.(\ref{eq:mf_eigeneq}) is given by
\begin{equation}
  \theta_k = \frac{\sqrt{\xi_k^2 + \eta_k \Delta^2}-\xi_k}
                     {\sqrt{\eta_k} \Delta}.
\end{equation}
Then, the mean field eigenvalue becomes
\begin{equation}
  E_{\sf mf} = -2 \sum_k s_k \sqrt{\xi_k^2 + \eta_k \Delta^2},
\end{equation}
and the expectation values become
\begin{eqnarray}
    \left \langle  S^z_k \right\rangle
      &=& -s_k  \frac{\xi_k}{\sqrt{\xi_k^2 + \eta_k \Delta^2}},
  \ \ \
    \left \langle  S^-_k \right\rangle
      = s_k  \frac{\sqrt{\eta_k}\Delta}{\sqrt{\xi_k^2 + \eta_k \Delta^2}},
   \label{eq:mfsmink}
  \\
    \left \langle  H_h \right\rangle
      &=&  - \sum_k s_k  \frac{\xi_k \eta_k}{\sqrt{\xi_k^2 + \eta_k \Delta^2}} - \frac{\Delta^2}{G}
  %\nonumber \\ & &
         - G \sum_k \frac{\eta_k s_k}{2} \left( 1-\frac{\xi_k}{\sqrt{\xi_k^2 + \eta_k \Delta^2}}\right)^2
\end{eqnarray}
Selfconsistency with the definition of $\Delta$ in Eq.(\ref{eq:mf_delta})
gives us the gap equation:
\begin{equation}
 \sum_k s_k  \frac{\eta_k}{\sqrt{\xi_k^2 + \eta_k \Delta^2}}
       = \frac{1}{G},
   \label{eq:mf_gap}
\end{equation}
while fixing the expectation value of $S^z$ gives us the number equation:
\begin{equation}
   -\sum_k s_k  \frac{\xi_k}{\sqrt{\xi_k^2 + \eta_k \Delta^2}} = M - L/2.
   \label{eq:mf_number1}
\end{equation}
In the thermodynamic limit $L\rightarrow \infty$, with $g = GL$ finite.
The expression for $\xi_k$ simplifies to:
\begin{equation}
  \xi_k =\eta_k/2 - \mu.
\end{equation}
Adding $1/2$ times Eq.(\ref{eq:mf_gap}) to Eq.(\ref{eq:mf_number1}) allows us
to rewrite the number equation as
\begin{equation}
  \frac{1}{L} \sum_k s_k  \frac{\mu}{\sqrt{\xi_k^2 + \eta_k \Delta^2}}
      = \frac{1}{2g} - 1/2 + \rho.
   \label{eq:mf_number2}
\end{equation}
We see that $\mu=0$ occurs at $g^{-1}= 1-2\rho$,
i.e. {\em special case (i)} from Sec.~\ref{sec:singularity}.
Another interesting situation occurs when
$\mu = \frac{\Delta^2}{2}$,
because then one has that 
$ \sqrt{\xi_k^2 + \eta_k \Delta^2} = \left| \eta_k/2 + \mu \right|$
and consequently $\theta_k=\Delta/k$.
By adding $1/2$ times Eq.(\ref{eq:mf_gap}) to Eq.(\ref{eq:mf_number2}) for this case,
one obtains (assuming all $\eta_k>0$) that
%\begin{equation}
%       \frac{1}{L} \sum_k s_k  \frac{ \eta_k/2 + \mu}
%                           {\sqrt{\xi_k^2 + \eta_k \Delta^2}}
%        = \frac{1}{2g} + \frac{1}{2g} - 1/2 + M/L,
%\end{equation}
%which simplifies to
$g^{-1}=1-\rho$,
i.e. {\em special case (ii)} from Sec.~\ref{sec:singularity}.

%------------------------------------------------------------------------------
%\bibliography{pips_library}% Produces the bibliography via BibTeX.
%------------------------------------------------------------------------------

\end{document}